\documentclass[conference,twocolumn]{IEEEtran}
\IEEEoverridecommandlockouts

\usepackage[utf8]{inputenc}
\usepackage[english]{babel}
\usepackage{graphics, graphicx}
\usepackage{amsmath, amsthm, amssymb, amsfonts}
\usepackage{algorithm}
\usepackage{algpseudocode}
\usepackage{url}
\usepackage{soul}  
\usepackage{tabularx}
\usepackage{comment}
\usepackage{soul}
\usepackage{subfig}
\usepackage{tikz}
\usepackage{cite}
\usepackage{xcolor}
\usepackage{glossaries}
\usepackage{tabu}
\usepackage{tabularx}
\usepackage[nocomma]{optidef}
\usepackage{soul}

\usepackage{lipsum}  
\usepackage{xcolor, colortbl}
\usepackage{mathtools, nccmath}
\usepackage{xpatch}
\xpatchcmd{\NCC@ignorepar}{%
\abovedisplayskip\abovedisplayshortskip}
{%
\abovedisplayskip\abovedisplayshortskip%
\belowdisplayskip\belowdisplayshortskip}
{}{}

\usepackage{hyperref}
\hypersetup{
    colorlinks=true,
    linkcolor=blue,
    filecolor=magenta,      
    urlcolor=cyan,
    pdftitle={Overleaf Example},
    pdfpagemode=FullScreen,
    }

\newacronym{awgn}{AWGN}{additive white Gaussian noise}
\newacronym{ao}{AO}{alternating optimization}
\newacronym{aod}{AoD}{angle-of-departure}
\newacronym{aoa}{AoA}{angle-of-arrival}
\newacronym{am}{AM}{amplitude modulation}
\newacronym{bs}{BS}{base station}
\newacronym{b5g}{B5G}{beyond 5th-Generation}
\newacronym{bf}{BF}{beamforming}
\newacronym{chest}{CHEST}{channel estimation}
\newacronym{csi}{CSI}{channel state information}
\newacronym{crb}{CRB}{Cramér-Rao bound}
\newacronym{dft}{DFT}{discrete Fourier transform}
\newacronym{dl}{DL}{downlink}
\newacronym{db}{DB}{double-bounce}
\newacronym{ee}{EE}{energy efficiency}
\newacronym{emf}{EMF}{electromagnetic field}
\newacronym[first= EMF exposure (EMFE)]{emfe}{EMFE}{electromagnetic field exposure}
\newacronym{fim}{FIM}{Fisher information matrix}
\newacronym[first = Genetic Algorithm (GA)]{ga}{GA}{genetic algorithm}
\newacronym[plural= STCMs, firstplural = space-time coding metasurfaces (STCMs)]{stcm}{STCM}{space-time coding metasurface}
\newacronym[first = Integrated Sensing and Communication]{isac}{ISAC}{integrated sensing and communication}
\newacronym{ifft}{IFFT}{inverse fast Fourier transform}
\newacronym{iot}{IoT}{Internet of Things}
\newacronym[plural = KPIs, firstplural = key performance indicators (KPIs)]{kpi}{KPI}{key performance indicator}
\newacronym{los}{LoS}{line-of-sight}
\newacronym{mimo}{MIMO}{multiple-input, multiple-output}
\newacronym{m-mimo}{M-MIMO}{massive MIMO}
\newacronym{mmwave}{mmWave}{millimeter-wave}
\newacronym[plural = MPCs, firstplural = multipath components (MPCs)]{mpc}{MPC}{multipath component}
\newacronym{mrt}{MRT}{maximum ratio transmission}
\newacronym{mcs}{MCS}{Monte Carlo simulations}
\newacronym{nue}{NUE}{non-intended UE}
\newacronym{pa}{PA}{power allocation}
\newacronym{pl}{PL}{path-loss}
\newacronym{pm}{PM}{phase modulation}
\newacronym{pdf}{PDF}{probability density function}
\newacronym{peb}{PEB}{position error bound}
\newacronym[plural = RISs, firstplural = reflective intelligent surfaces (RISs)]{ris}{RIS}{reflective intelligent surface}
\newacronym{ra}{RA}{resource allocation}
\newacronym{rv}{RV}{response vector}
\newacronym{rcs}{RCS}{radar cross section}
\newacronym{se}{SE}{spectral efficiency}
\newacronym{snr}{SNR}{signal-to-noise ratio}
\newacronym[plural = SPs, firstplural = scatter points (SPs)]{sp}{SP}{scatter point}
\newacronym{scs}{SCS}{sub-carrier spacing}
\newacronym{sb}{SB}{single-bounce}
\newacronym{tdd}{TDD}{time division duplex}
\newacronym{toa}{ToA}{time-of-arrival}
\newacronym[first = Uplink (UL)]{ul}{UL}{uplink}
\newacronym{ula}{ULA}{uniform linear array}
\newacronym{upa}{UPA}{uniform planar array}
\newacronym{ura}{URA}{uniform rectangular array}
\newacronym[plural = UEs, firstplural = user equipments (UEs)]{ue}{UE}{user equipment}
\newacronym{5g}{5G}{5th-Generation}
\newacronym{6g}{6G}{6th-Generation}
\newacronym{3d}{3D}{three-dimensional}
\newacronym{2d}{2D}{two-dimensional}
\definecolor{cian}{rgb}{.02,.7,.95}
\definecolor{gold}{rgb}{0.85,.66,0}
\definecolor{Gray}{gray}{0.95}

\newcommand{\bs}[1]{\boldsymbol{#1}}

\newcommand{\bvp}{\boldsymbol{\varphi}}
\newcommand{\bab}{{{\bf a}_\rB}}
\newcommand{\babt}{{{\bf a}^\rT_\rB}}
\newcommand{\baR}{{{\bf a}_\rR}}
\newcommand{\baRt}{{{\bf a}^\rT_\rR}}

\newcommand{\rB}{{\rm B}}
\newcommand{\rR}{{\rm R}}
\newcommand{\rS}{{\rm S}}
\newcommand{\rT}{{\top}}

\DeclareMathOperator*{\argmax}{arg\,\max}

\title{{Assessing the Potential of Space-Time-Coding Metasurfaces for Sensing and Localization}
\thanks{
This work was supported in part by the Coordenação de Aperfeiçoamento de Pessoal de Nível Superior - Brasil (CAPES) – PDSE Program and by the National Council for Scientific and Technological Development (CNPq) of Brazil under Grants 405301/2021-9, 141485/2020-5, and 310681/2019-7. The work of M. V. Vejling and P. Popovski were supported in part by the Villum Investigator Grant “WATER” financed by the VILLUM Foundation, Denmark.
}
}
\author{
\IEEEauthorblockN{
    {
    {Herman L. dos Santos}}\IEEEauthorrefmark{1},
    {
    {Martin Voigt Vejling}}\IEEEauthorrefmark{2}\IEEEauthorrefmark{3},
    {
    {Taufik Abrão}}\IEEEauthorrefmark{1},
    and {
    {Petar Popovski}}\IEEEauthorrefmark{2} \\
        }
        \IEEEauthorblockA{
        \IEEEauthorrefmark{1}\textit{\small Department of Electrical Engineering, Universidade Estadual de Londrina}, Londrina, Brazil\\
        \IEEEauthorrefmark{2}\textit{\small Department of Electronic Systems, Aalborg University}, Aalborg, Denmark\\
        \IEEEauthorrefmark{3}\textit{\small Department of Mathematical Sciences, Aalborg University}, Aalborg, Denmark\\
        \small E-mail: hermanlds@gmail.com, mvv@es.aau.dk, taufik@uel.br, and petarp@es.aau.dk 
        }
}
\begin{document}
\maketitle

\thispagestyle{plain}
\pagestyle{plain}

\begin{abstract}
Intelligent metasurfaces are one of the favorite technologies for integrating sixth-generation (6G) networks, especially the reconfigurable intelligent surface (RIS) that has been extensively researched in various applications. {In this context, a feature that deserves further exploration is} the frequency scattering {occurs} when the elements are periodically switched, {referred to as} Space-Time-Coding metasurface (STCM) topology. This type of topology {causes impairments to the established communication methods} by generating undesirable interference both in frequency and space, which is worsened when using wideband signals. {Nevertheless, it has the potential to bring forward useful features for sensing and localization.} This work exploits STCM sensing capabilities in target detection, localization, and classification using narrowband downlink pilot signals at the base station (BS). The results of this novel approach reveal the ability to retrieve a scattering point (SP) localization within the sub-centimeter and sub-decimeter accuracy depending on the SP position in space. We also analyze the {associated} detection and classification probabilities, which show reliable detection performance in the whole analyzed environment. In contrast, the classification is bounded by physical constraints, and we conclude that this method presents a promising approach for future integrated sensing and communications (ISAC) protocols by providing a tool to perform sensing and localization services using legacy communication signals.
\end{abstract}

\section{Introduction}
\gls{isac} is a reality and key enabler for \gls{b5g} and \gls{6g} technologies \cite{FangX2023}. Sensing applications are of great interest in smart transportation \cite{ChengX2022}, \gls{iot} \cite{LinaM2023}, green communications \cite{YingDu2023}, and surveillance \cite{ZhangM2023,SunB2022}. \gls{isac} finds motivation in spectrum scarcity since both sensing/radar and communication functions have ever-increasing \glspl{kpi} as the generations of telecommunication evolve and their frequency bands usage increase to the point of overlapping, causing mutual interference \cite{JunshengM2023} between the two system's {dedicated} signals.

{\gls{b5g} systems target} localization-oriented applications, such as vehicular interaction with the environment, automatic robots in manufacturing sharing space with humans, environmental monitoring, crowd and drone monitoring, human-machine interaction with gesture and activity recognition, and smart home monitoring \cite{LiuFan2022}. All these applications demand knowledge of environment configuration{, performed under spectrum limitations, as it may not be possible to have dedicated units for communication and sensing with dedicated frequency bands.}

We highlight the \gls{m-mimo} hardware deployment, which enables the improvement of both sensing and communication metrics and the emerging metasurfaces. These devices can be classified as linear, commonly known as \gls{ris} \cite{WangX2023}, and non-linear metasurfaces.  Non-linear metasurfaces can be either passive or active. Passive metasurfaces with non-linear topology, inserting an extra-degree of freedom into \gls{ris} classification, enabling manipulation of frequency-momentum spaces, while active antennas metasurfaces are capable of manipulating every fundamental property of electromagnetic waves \cite{WuGB2023}. A specific case of non-linear passive metasurface is the \gls{stcm} \cite{Zhang2018}, which consists of a topology that is employed as a reflector and space-frequency scatterer, bringing a unique and valuable property for sensing and localization services.

\gls{stcm}, also named time-domain digital coding metasurface \cite{DaiJunYan2020TDCM}, is a class of low-cost non-linear metasurfaces that manipulate electromagnetic waves in both space and frequency domains. It can be viewed as a \gls{ris} topology that exploits not only space scattering \cite{Zhang2018}, \textit{i.e.} fixing the coding sequences in time, but also multiple frequencies spread in the space. The periodic switching of \gls{stcm} elements induces desired pre-programmed harmonic frequencies over space, generating power-scattered signals
at different frequencies that propagate to different directions instead of the controlled reflection only. Since the harmonic frequencies are directly related to space (direction), this can be used in sensing and localization for retrieving environmental information about \glspl{sp}. 

To the best of the authors' knowledge, the \gls{stcm} have not been sufficiently explored in the context of sensing {and localization}. Research has been conducted in applying such topology for communication applications using narrowband signals \cite{Robin2021_fMIRS}. The authors explored the multipath channel decoupling, and some gains are observed in the channel {response}, channel estimation, and spectral efficiency. Performing this decoupling demands knowledge of the surrounding environment to at least estimate the {propagated paths of the signal. } 
Hence, to achieve maximum communication performance, a sensing step to retrieve the propagation scenario is highly desirable. The paper's contribution is aligned with the communication system framework presented in \cite{Robin2021_fMIRS}; hence, we provide an architecture for \gls{isac} by {integrating both sensing and localization procedures} in the usage of the \gls{stcm}. {On the other hand, \cite{Robin2021_fMIRS} deals mainly with communication in the STCM context, whereas the current contribution deals chiefly with sensing and localization aspects. Hence, the previous work \cite{Robin2021_fMIRS} is complementary, such that in the future both works can be combined in integrated sensing and communication systems.}

\vspace{2mm}
\noindent{\textbf{\textit{Contributions}}: In this paper, we evaluate the capabilities of \gls{stcm} in \gls{m-mimo} \gls{isac} for future generation of telecommunications. Specifically, we deploy a framework consisting of exploiting \gls{dl} pilots, hence generating no further communication overhead, and the echo signals at the \gls{bs} to perform tasks such as target detection and classification. We {derive and} compare theoretical bounds for both detection and classification procedures and analyze the system performance with a \gls{ris}-assisted system and a coordinated double \gls{bs} system. The results related to \gls{stcm} present a major advantage over the conventional \gls{ris}-assisted system, which can not solely obtain the localization of \gls{sp} using this minimal signaling, but it
does not need coordination or backhaul signaling, as in double \gls{bs} scheme, since the processing can be done centralized at the \gls{bs}.}

The remainder of this paper is organized as follows. Section \ref{sec:II} describes the ISAC M-MIMO system model. The space-time sensing with \gls{stcm} is treated in Section \ref{sec:III}, where we derive theoretic Cramer-Rao {lower bound 
information for estimation of the channel parameters, and for localization of \glspl{sp}}. Section \ref{sec:IV} presents classification bounds for differentiating 
{distinct}  \glspl{sp} {types}. {Numerical illustration is provided} in Section \ref{sec:V}, including the evaluation of the system capabilities. The main conclusions are summarized in Section \ref{sec:VI}.

\section{System Model}\label{sec:II}
Consider an \gls{isac} \gls{m-mimo} system where a full-duplex \gls{bs} equipped with $M$ antennas, {arranged} in a \gls{ula}, communicates with $K$ {single-antenna} \glspl{ue} in \gls{dl} using narrowband signals, and is supported by an $N = N_y\times {N_x}$ voltage controlled PIN-diode \gls{stcm} {panel}. The voltages of \gls{stcm} elements are individually controlled {by} the \gls{bs} side, periodically switching their phase-shift profile with period $T_0$. The periodic switching alters the frequency-momentum space, and equivalent phase shifts are produced in the spectrum around the central frequency $f_c$ at the switching frequency $f_0={T_0}^{-1}$ and its' harmonics.
Hence, {this topology of usage} induces controlled space scattered harmonic frequencies at $f_c \pm mf_0$, for $m= 0, \pm 1,\dots, \pm m_f$ and $m_f > 0$. These harmonic frequencies are spread to the whole spectrum but more attenuated as $m$ increases. Considering an incident wave with azimuth and elevation \gls{aoa} $\bvp_A = [\phi_A,\theta_A]$, respectively; the time-domain far-field pattern evaluated at \gls{aod} $\bvp_D = [\phi_D,\theta_D]$, as in \cite{Zhang2018}, is given by:
\begin{equation}
    \begin{split}
        f(\bvp_D,\bvp_A,t) &= \sum\limits_{q=1}^{N_y}\sum\limits_{p=1}^{N_x} E_{pq}(\bvp_D) \Gamma_{pq}(t)\\
        &\quad \times \exp\Big\{j\big({\bf k}_{\lambda_c}^\rT(\bvp_D) + {\bf k}_{\lambda_c}^\rT(\bvp_A)\big) 
        {\bf q}_{pq}\Big\}
    \end{split}
\end{equation}
where $E_{pq}(\bvp_D)$ denotes the far-field wave pattern of the $(p,q)$-th element observed at $\bvp_D$, 
${\bf q}_{pq} = [q_{pq,x},\,q_{pq,y}, \, \,q_{pq,z}] \in \mathbb{R}^{3\times 1}$ is the position of the $(p, q)$-th element of the \gls{stcm} panel in local coordinates, ${\bf k}_{\lambda_c}(\bvp) = \frac{2\pi}{\lambda_c} [\cos(\phi)\sin(\theta), \sin(\phi)\sin(\theta), \cos(\theta)]^\rT$ is the wavenumber vector for wavelength of the carrier frequency $\lambda_c=\frac{c}{f_c}$, $c$ is the speed of light,
 and $\Gamma_{pq}(t)$ is the function defining the periodic time-modulated reflection coefficients of the {metasurface elements}:
\begin{equation}
    \Gamma_{pq}(t) = \sum_{\ell=1}^L\Gamma^\ell_{pq}U^\ell_{pq}(t), \quad (0<t<T_0),
    \label{eq:Gamma(t)}
\end{equation}
\noindent being $U^\ell_{pq}(t)$ a pulse function with period $T_0$ and $L$ is the time coding-sequence length. $\Gamma^\ell_{pq}$ assumes values $\{0,1\}$ in \gls{am} scheme, or $\{-1,1\}$ for \gls{pm}. The frequency-domain far-field pattern of {the} $m$-th harmonic $\eta_m \in \mathbb{C}$ is 
{described} as\footnote{We suppressed the Fourier transform evaluation, which can be further analyzed in \cite{Zhang2018}.} 
%
%
\begin{equation}\label{eq:STCMFm}
    \begin{split}
        \eta_m(\bvp_D,\bvp_A) &= \sum\limits_{q=1}^{N_y}\sum\limits_{p=1}^{N_x} E_{pq}(\bvp_D) a^m_{pq}\\
        &\qquad \times \exp\Big\{j \big({\bf k}_{\lambda_m}^{\rT}(\bvp_D) + {\bf k}_{\lambda_m}^{\rT}(\bvp_A)\big) {\bf q}_{pq}\Big\}
    \end{split}
\end{equation}
\noindent where $a^m_{pq}$ is the Fourier-series coefficients of the periodic function $\Gamma_{pq}(t)$, Eq. \eqref{eq:Gamma(t)}, defined as
\begin{equation}\label{eq:aFScoeff}
a_{pq}^m = \sum\limits^L_{\ell=1} \frac{\Gamma_{pq}^\ell}{L}{\rm sinc}\left(\frac{\pi m}{L}\right)\exp\left(\frac{-j\pi m (2\ell-1)}{L}\right).
\end{equation}

By controlling the switching variable $\Gamma^\ell_{pq}$ it is possible to obtain specific scattering patterns in each of the $m$-th harmonics. {For simplicity,} in \gls{pm} scheme, setting an element as $-1~(1)$ means setting the output phase alignment as $-\pi ~(\pi)$ and the combination during the $L$ length time-coding sequence generates phase alignment in the whole degree range, thus specific beam alignment can be performed by calculating the \gls{ifft}. 

\subsection{Sensing via Metasurface space-frequency features} 
\begin{figure}[!htbp]
\centering
\subfloat[\gls{stcm}-assisted \gls{m-mimo} system containing 1 \gls{ue}, 1 \gls{nue}, and 1 Obj. The lines (solid, dashed, dotted) represent the power gain of the signal after reflection due to the \gls{sp} \gls{rcs} $\sigma_i$.] 
{\includegraphics[trim = 5mm 1mm 6mm 0mm, width=1\linewidth]{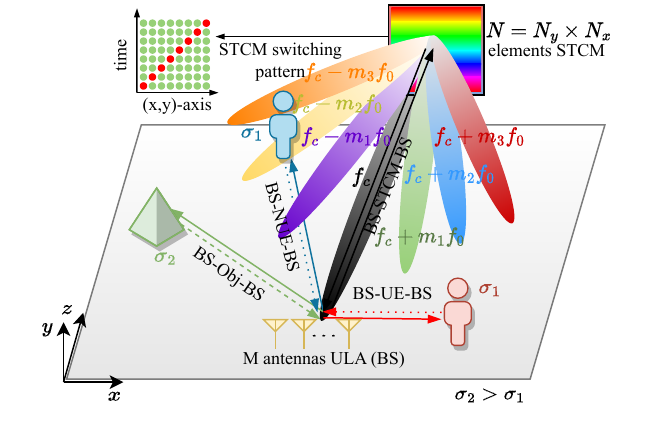}\label{fig:system}}\\
\vspace{-1mm}
\subfloat[SB (\textit{c1.} and \textit{c2.} terms) and DB (\textit{c3.} and \textit{c4.} ) signals with frequency modulation around central frequency $f_c$.]{\includegraphics[trim = 9mm 7mm 15mm 0mm, width=.98\linewidth]{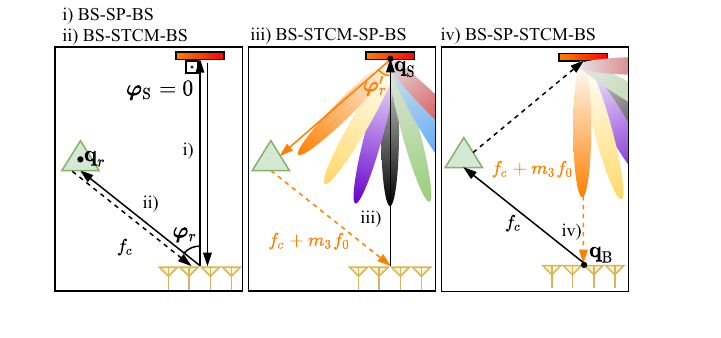}\label{fig:bouncing}}

\subfloat[Single-emitter \gls{stcm}-aided \gls{isac} \gls{sp} localization by deploying trigonometric relationship.]
{\includegraphics[trim= 5mm 4.5mm 14mm 0mm, clip, width = .99\linewidth]{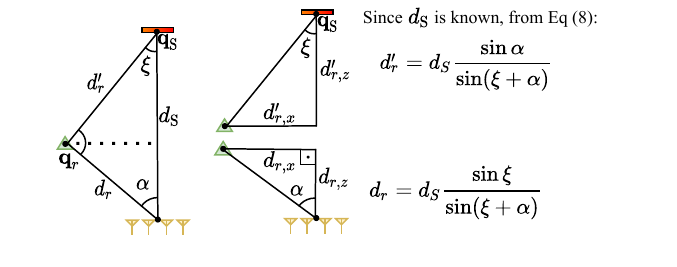}\label{fig:toyExample}}

\vspace{-.1mm}
\caption{\small Scenario and bouncing diagram, indicating the BS-STCM ($\bvp_{\rS}$) angle, the \gls{bs}-\gls{sp} angle ($\bvp_{r}$), and \gls{stcm}-\gls{sp} angle ($\bvp_{r}'$).}
\label{fig:downlink}
\end{figure}

We employ the space-frequency characteristic brought by the metasurface to perform sensing relying on \gls{dl} pilot signals, thus integrating sensing {becomes} 
a mandatory step in communication protocols. {During the pilots' transmission, the \gls{stcm} is kept with a fixed and pre-defined switching pattern (See Fig. \ref{fig:bouncing}), thus working \textit{independently} of the \gls{bs} and not demanding control channel resources, which can hugely affect the 
{time sensing} \cite{saggese2023_impactRISTiming}.} Upon transmission, {the} \gls{bs} receives the backscattered echo signal and retrieves environment information from the reflections in \glspl{sp}. We consider two types of \glspl{sp}: static environment objects, namely Obj, and humans, \glspl{nue} and \glspl{ue}, both modeled as point-targets. The main difference between these \glspl{sp} is that the human bodies absorb a higher parcel of the impinging signal, thus having a lower reflection coefficient, while the objects reflect with higher power gain. Fig. \ref{fig:downlink} {depicts} a comprehensive diagram supported by Eq. \eqref{eq:echo} nomenclature.

We investigate a scenario containing $K$ \glspl{ue}, $\bar{K}$ \glspl{nue}, and $O$ Objs. The set of environment \glspl{sp} is defined as \resizebox{1\linewidth}{!}{$\mathcal{R} = \{{\rm UE}_1,\dots,{\rm UE}_K , {\rm NUE}_1, \dots, {\rm NUE}_{\tilde{K}}, {\rm Obj}_1, \dots, {\rm Obj}_{O} \}$}, and $|\mathcal{R}| = (K + \tilde{K} + O)$. Fig. \ref{fig:downlink} presents the interaction of the elements in \gls{dl} \gls{stcm}-based
\gls{m-mimo} pilot transmission. We remark that \gls{bs} and \gls{stcm} share the same normal, \textit{i.e.} are disposed with the horizontal centers' aligned to the origin. Upon transmission, the signals reflect in every \gls{sp} and in the \gls{stcm} and are received back at the \gls{bs}, characterizing the \gls{sb} components: \textit{c1)} \gls{bs}-\gls{stcm}-\gls{bs} and \textit{c2)} \gls{bs}-\gls{sp}-\gls{bs}, as presented in Fig \ref{fig:bouncing}. Moreover, the signals are subject to double-bouncing paths, {\it i.e.}, \textit{c3)} \gls{bs}-\gls{stcm}-\gls{sp}-\gls{bs} and \textit{c4)} \gls{bs}-\gls{sp}-\gls{stcm}-\gls{bs}, which are, respectively, replicas of the transmitted signal reflected from the \gls{stcm} towards the \gls{sp}, and from the \gls{sp} towards the \gls{stcm}, and both reflected back to the \gls{bs}, exemplary illustrated in Fig. \ref{fig:bouncing}. The received echo signal in {the} $m$-th harmonic component ${\bf Y}^{\rm echo}_m \in \mathbb{C}^{M\times {S}}$ is formally described as\footnote{{\it c2)} is zero for $m\neq 0$. {Expression adapted from \cite{kim2022}.}}:
\begin{equation}\label{eq:echo}
\begin{split}
     &{\bf Y}^{\rm echo}_m  \triangleq   \bigg(\underbrace{\bar{\beta}_{\rS} \eta_m(\bvp_{\rS},\bvp_{\rS})\bab(\bvp_{\rS})\babt(\bvp_{\rS})d_{{m}}({\tau_\rS})}_{\text{{\it c1)} BS-STCM-BS}}\\
     &\:+ \sum\limits_{r\in\mathcal{R}} \underbrace{\bar{\beta}_{r} \bab(\bvp_{r})\babt(\bvp_r)d_{{m}}({\tau_r})}_{\text{{\it c2)} BS-SP-BS}} \\
     &\:+ \sum\limits_{r\in\mathcal{R}} \underbrace{\bar{\bar{\beta}}_r \eta_m(\bvp_{r}',\bvp_{\rS})\bab(\bvp_{r})\babt(\bvp_{\rS})d_{{m}}({\tau}_{r})}_{\text{{\it c3)} BS-STCM-SP-BS}}\\
     &\:+\sum\limits_{r\in\mathcal{R}} \underbrace{\bar{\bar{\beta}}_{r} \eta_m(\bvp_\rS,\bvp_r')\bab(\bvp_{\rS})\babt (\bvp_{r})d_{{m}}({\tau}_{r})}_{\text{{\it c4)} BS-SP-STCM-BS}}\bigg)    
    {\bf X} + {\bf N}_m,
\end{split}
\end{equation}
\noindent where the scalars $\{\bar{\beta}_{r},\bar{\bar{\beta}}_{r}\} \in \mathbb{C}^+$ encompasses the path gain between the $r$-th \gls{sp} and the \gls{bs} in \gls{sb} and \gls{db}, respectively. 
$\bvp_r$, $\bvp_r'$, {and $\bvp_{\rS}$} are the \gls{bs}-\gls{sp}, \gls{stcm}-\gls{sp}, and \gls{bs}-\gls{stcm} angles, respectively, $\bab(\bvp_r) \in \mathbb{C}^{M\times 1}$ 
\cite[Eq. \eqref{eq:steervec}]{keykhosravi2022}, is the array \gls{rv} at the \gls{bs} defined by the wavenumber ${\bf k}$ and antenna elements positions {matrix} {${\bf Q}_\rB =[\mathbf{q}_0, \mathbf{q}_1, \dots \mathbf{q}_{M}] \in \mathbb{R}^{M\times 3}$ w.r.t. $\bvp$,
\begin{equation}\label{eq:steervec}
{\bf a}_{\rm B}(\bvp)={\rm exp}\big(j\mathbf{Q}_{\rm B} \mathbf{k}_{\lambda_c}(\bvp)\big),
\end{equation}}%
$d_{{m}}(\tau) \triangleq {\rm exp}(-j2\pi\tau mf_0)$ is the Doppler phase shift dependent on the 
\glspl{sp} {positions ${\bf q}_r$ and relative} velocity {in local coordinates ${\bf v} \in \mathbb{C}^{3\times 1}$, i.e. $\tau = 2\frac{|{\bf q}_\rB - {\bf q}_r|^\top}{\|{\bf q}_\rB - {\bf q}_r\|^2 \lambda} {\bf v}$} \cite{Wymeersch2022:Fundamentals}. Throughout this work, we have adopted fully static scenarios.
Hence, ${\tau_r=\tau_\rS}=0$, so $d_{{m}}(\tau) = 1$.
${\bf X} = {\bf w}{\bf s}\,  \in \mathbb{C}^{M\times S}$ denotes the $S$ transmitted symbols {vector } ${\bf s} \in \mathbb{C}^{1\times S}$ using a precoding vector ${\bf w}\in \mathbb{C}^{M\times 1}$, encompassing both directivity and transmit power, ${\bf N}_m \sim \mathcal{CN}({\bf 0}, \sigma_n^2{\bf I}_M)$ is the \gls{awgn} at the \gls{bs} and $\eta_m(\bvp_{D},\bvp_{A})\in \mathbb{C}$, Eq. \eqref{eq:STCMFm}, denotes the \gls{stcm} steered \gls{rv} w.r.t. \gls{aoa} $\bvp_{\rm A}$ and \gls{aod} $\bvp_{\rm D}$ at $m$-th harmonic.
Finally, the \gls{sb} and \gls{db} path gains are calculated, respectively, as 
\begin{subequations}
\begin{align}
\bar{\beta}_r =  G(\bar{d}_r)\exp\left(-j2\pi \frac{\bar{d}_r}{c}\right)\sigma_r\nu, \label{eq:pathSB} \\
\bar{\bar{\beta}}_r = G(\bar{\bar{d}}_r) \exp\left(-j2\pi \frac{\bar{\bar{{d}}}_r}{c}\right)\sigma_r \nu,  \label{eq:pathDB}\\
G(d) = \frac{\sqrt{E_s}\lambda}{4\pi d^\iota}
\end{align}
\end{subequations}
\noindent for $d>0$, and where
$\bar{d}_r = 2d_r$ and $\bar{\bar{d}}_r = d_\rS+d_r+{d_{r'}}$, {with} $d_r=\|{\bf q}_\rB - {\bf q}_r\|$, $d_\rS = \|{\bf q}_\rB - {\bf q}_\rS\|$, and $d_{r'} = \|{\bf q}_\rS - {\bf q}_r\|$ {being} the \gls{bs}-\gls{sp}, \gls{bs}-\gls{stcm}, and \gls{stcm}-\gls{sp} Euclidean distances, respectively, with ${\bf q}_{\rm B}$, ${\bf q}_{\rm S}$, and ${\bf q}_r$ being the coordinates of the center of the \gls{bs} antennas, center of the \gls{stcm} elements, and $r$-th \gls{sp} coordinate, respectively. $\iota$ encompasses the square root path-loss exponent 
$\bar{d}_r/c$, and  $ \bar{\bar{d}}_r/c$ are the delays of \gls{sb} and \gls{db} echo signals due to different \gls{toa}, $E_s$ is the symbol energy. The parameter $\sigma^{2}_r {\in \mathbb{R}^+}$ represents the \gls{rcs} of the $r$-th \gls{sp} which models both the attenuation and phase shifts on backscattered signals and is dependent on object properties, incident angle, carrier frequency, and its effective reflection area \cite{Jarvis2021}; and $\nu \sim \mathcal{CN}(0,\sigma^2_\nu)$ models multipath components related to specular reflection, diffraction, refraction, and creeping wave phenomenons \cite{Tien2017HumanRCS} around \glspl{sp}.

The set \gls{bs}-\gls{sp}-\gls{stcm} form a triangle in space, Fig. \ref{fig:toyExample}, with its' side lengths as $d_r$ (\gls{bs}-\gls{sp}), $d_\rS$ (\gls{bs}-\gls{stcm}), and $d_r'$ (\gls{stcm}-\gls{sp}), which are \textit{a priori} unknown. However, it is feasible to assume a fixed position for the \gls{stcm}, since the control of such a device is done by a backhaul link to the \gls{bs}; as a result, one can derive a trigonometric relationship between these three spatial points. Assuming a \gls{2d} plane defined by the localization of \gls{bs}, \gls{sp}, and \gls{stcm} sharing the same vertical $y$-axis height, simplifying $\bvp_r = \alpha$ and $\bvp_r' = \xi$, regarding only the azimuth angles $(\theta_r = 0 \ \forall \ r)$, the trigonometric relationship simplifies: 
\begin{equation}\label{eq:angRelation}
    \frac{d_\rS}{\sin \zeta} = \frac{d_r}{\sin {\xi}} = \frac{d_r'}{\sin {\alpha}},
\end{equation}
where $\zeta = \pi - {\alpha} - {\xi}$, giving us enough information to localize an \gls{sp}, avoiding to calculate the distance from a \gls{sp} by the path gain. Finally, 
one can calculate $d_r= d_S \frac{\sin\xi}{\sin(\alpha+{\xi})}$, with the $r$-th \gls{sp} coordinates ${\bf q}_r = {[q_{r,x},q_{r,y},q_{r,z}]}
$ as $q_{r,x} = q_{\rB,x} + d_r\sin\alpha$ and $q_{r,z} = q_{\rB,z}+d_r\cos\alpha$, while $q_{r,z} = 0$.

\section{Space-frequency sensing with STCM}\label{sec:III}

The basic framework for recovering an \gls{sp} localization in the proposed \gls{stcm}-aided \gls{isac} system relies on using trigonometrical properties of the multipath bouncing, accessing the potential of recovering them at the \gls{bs} upon receiving the echo signal. When the \gls{bs} can recover the angles between itself and the \gls{sp} and \gls{stcm}-\gls{sp}, it provides {a robust} estimation method {regarding} the stochastic characteristics of the path. Classical triangulation relies on synchronization and sensing by multiple \glspl{bs}, which is demanding both computationally and in terms of energy consumption. Alternatively, the \gls{stcm} acts like a semi-passive {\it artificial} signal emitter, and its' {intrinsic} properties are useful for estimating the angle between this device and potential \glspl{sp}. Thus, the processing is done centralized at the \gls{bs} by estimating the \glspl{sp}' \gls{aoa} and the angle between the \glspl{sp} and the \gls{stcm}. For the sake of simplicity, throughout this work we consider:  a) 2-D scenarios, as introduced in Sec. \ref{sec:II}, and b) non-mobility, static scenarios, hence $\{\bar{\tau},\bar{\bar{\tau}}\} =0$. Fig. \ref{fig:toyExample} depicts a comprehensive diagram justifying the trigonometric calculation in Eq.\eqref{eq:angRelation}.

The estimation of angle $\alpha$ is immediate by any classic estimator, such as MUSIC, ESPRIT, MVDR, and others, while estimating $\xi$ relies upon the relationship between the received power gain of harmonics frequencies generated by \gls{stcm}. Knowing the distance $d_\rS$, we can estimate $d_r$, hence retrieving the \gls{sp} location with high precision, using the 
trigonometric relation in Eq. \eqref{eq:angRelation}, {Fig. \ref{fig:toyExample}.}

\vspace{2mm}
\noindent{\textbf{\textit{Assumptions}}}. 
{The following assumptions are {adopted}}. The components {\it c1), c2)}, and {\it c3)-c4)} {in Eq. \eqref{eq:echo}} have different \gls{toa} at the \gls{bs} array, {hence} the {signal} processing {can be} {carried out} separately. We divide the processing framework by using {\it c2)} to estimate $\alpha$ and {\it c3)-c4)} to estimate $\xi$\footnote{Note that this is a simplification to evaluate the capabilities with this novel approach.} by {deploying} the relationship between the harmonics power level {generated at the \gls{stcm} and considering} far-field {condition, in same way as adopted} in \cite{Zhang2018}. In other words, different \gls{aod} and \gls{aoa} at the \gls{stcm} generate a unique vector of power gains of components observed {at} the \gls{bs}, which are a basis of inference {for} this angle. The latter components which {experience} \gls{db}, can not be dissociated since they have the same \gls{toa} due to their traveled distance being the same. Hence, we estimate angle $\xi$ by deploying harmonic signals generated at the \gls{stcm} arriving with the different power gains.

\subsection{{Localization Problem}}

Under these considerations, to formulate and assess the \gls{sp} {\it localization} the capability of such a setup, we rely on the information theory \gls{crb} computations. 
We also assume the position and orientation of the STCM are known in advance. Indeed, such an assumption is reasonable since the estimation can be done in a prior step where one might use time-orthogonal STCM sequences to cancel any other interference. Besides, the BS knows the STCM position since its control relies on a backhaul link; hence, such prior information is feasible. 

\vspace{2mm}
\noindent\textbf{\textit{Cramér-Rao Bound}}: The \gls{crb} gives us the lower bound of the covariance among unbiased estimators on unknown deterministic and/or stochastic parameters. Estimators achieving this lower bound are called unbiased and {fully} efficient estimators \cite{FontanelliD2021}. Such a property can be found, {\it e.g.}, in the maximum likelihood estimator which is unbiased and efficient in the context of linear models, and for non-linear models, this weakens to asymptotically unbiased and efficient. To estimate the localization of an \gls{sp} one can 
estimate the sufficient information $\alpha$ and $\xi$ angles, Fig. \ref{fig:toyExample}, characterizing a multivariate case, where the quality of estimation of both angles subject to the path gains. Relying on the capacity of processing separately the \gls{sb} and \gls{db} components, we split the respective parameter sets to be estimated as:
\begin{equation}
    \begin{split}
        \boldsymbol{\Psi}^{\rm SB} = [\alpha,\tilde{\bar{\boldsymbol{\beta}}}] =  [\alpha_1,\dots,\alpha_{|\mathcal{R}|},{\rm Re}\{\bar{\beta}_1\}, {\rm Im}\{\bar{\beta}_1\},\\ \dots, {\rm Re}\{\bar{\beta}_{|\mathcal{R}|}\}, {\rm Im}\{\bar{\beta}_{|\mathcal{R}|}\}] \in \mathbb{R}^{1\times3|\mathcal{R}|} ,
    \end{split}
\end{equation}
and  
\begin{equation}
\begin{split}
    \boldsymbol{\Psi}^{\rm DB} =[\xi,\tilde{\bar{\bar{\boldsymbol{\beta}}}}] = [\xi_1,\dots,\xi_{|\mathcal{R}|},{\rm Re}\{\bar{\bar{\beta}}_1\}, {\rm Im}\{\bar{\bar{\beta}}_1\},\\ \dots, {\rm Re}\{\bar{\bar{\beta}}_{|\mathcal{R}|}\}, {\rm Im}\{\bar{\bar{{\beta}}}_{|\mathcal{R}|}\}] \in \mathbb{R}^{1\times3|\mathcal{R}|}, 
\end{split}
\end{equation}
which are the deterministic parameters ($\alpha,\xi$) and mixed deterministic-stochastic $(\bar{\beta},\bar{\bar{\beta}})$ organized as $\tilde{\bar{{\boldsymbol{\beta}}}} = [{\rm Re} \{\bar{\boldsymbol{\beta}}\}, {\rm Im} \{\bar{\boldsymbol{\beta}}\}]$ and $\tilde{\bar{\bar{\boldsymbol{\beta}}}} = [{\rm Re} \{\bar{\bar{\boldsymbol{\beta}}}\}, {\rm Im} \{\bar{\bar{\boldsymbol{\beta}}}\}]$ unknown variables. 

Let $\mathcal{M} = \{-m_f, -(m_f-1), \dots, m_f\}$ be the set of analyzed harmonics, including 0 (central frequency), and $|\mathcal{M}|$ the cardinality of the set.
{Before proceeding, we rearrange the echo signal \eqref{eq:echo} into the \gls{sb} and \gls{db} parts while stacking the signals in a vector. Let $\hat{\bf n}^{\rm SB}={\rm vec}({\bf N}_{0})\in \mathbb{C}^{\cdot M \cdot S\times 1}$ and $\hat{\bf n}^{\rm DB} = [{\rm vec}({\bf N}_{-m_f}), \dots, {\rm vec}({\bf N}_{m_f})]^\top \in \mathbb{C}^{|\mathcal{M}| \cdot M \cdot S\times 1}$, where vec$(\cdot)$ refers to the vectorization of the matrix, be the vectorized noise and let
\begin{equation*}
    {\bf H}^{\rm SB}({\bf q}_r) = {\rm vec}(\bab(\bvp_{r})\babt(\bvp_r){\bf X})^{\rT} \in \mathbb{C}^{M \cdot S\times 1}
\end{equation*}
be the deterministic part of the \gls{sb} signal, {\it i.e.} component \textit{c2)}. Then, the \gls{sb} signal is given as
\begin{equation}\label{eq:sb_signal}
    \hat{\bf y}^{\rm SB} = \sum\limits_{r\in\mathcal{R}} \hat{\bf u}^{\rm SB}({\bf q}_r) + \hat{\bf n}^{\rm SB}
\end{equation}
where $\hat{\bf u}^{\rm SB}({\bf q}_r) = \bar{\beta}_{r} {\bf H}^{\rm SB}({\bf q}_r)$. Similarly, for the \gls{db} signal let us define ${\boldsymbol \eta}(\bvp, \bvp') = [{\bf \eta}_{-m_f}(\bvp, \bvp'), \dots, {\bf \eta}_{m_f}(\bvp, \bvp')]^\rT$, and let
\begin{equation*}
    \begin{split}
        &{\bf H}^{\rm DB}({\bf q}_r) = {\boldsymbol \eta}(\bvp_r', \bvp_{\rm S}) \otimes {\rm vec}(\bab(\bvp_{r})\babt(\bvp_{\rm S}){\bf X})^{\rT} \\
        &\quad + {\boldsymbol \eta}(\bvp_{\rm S}, \bvp_r') \otimes {\rm vec}(\bab(\bvp_{\rm S})\babt(\bvp_{r}){\bf X})^{\rT} \in \mathbb{C}^{|\mathcal{M}| \cdot M \cdot S\times 1}.
    \end{split}
\end{equation*}
where $\otimes$ denotes the Kronecker product.
The \gls{db} signal is then
\begin{equation}\label{eq:db_signal}
    \hat{\bf y}^{\rm DB} = \sum\limits_{r\in\mathcal{R}} \hat{\bf u}^{\rm DB}({\bf q}_r) + \hat{\bf n}^{\rm DB}
\end{equation}
where $\hat{\bf u}^{\rm DB}({\bf q}_r) = \bar{\bar{\beta}}_{r} {\bf H}^{\rm DB}({\bf q}_r)$.}%

We use this arrangement to calculate the mutual information matrices of \gls{sb} and \gls{db} paths. In the present case of \gls{awgn}, the \gls{fim} is computed as
\begin{equation}\label{eq:fim}
\begin{split}
    \hat{\bf{F}}_{i,j} = \frac{2}{\sigma_{n}^2}{\rm Re}\left\{\frac{\partial}{\partial {\Psi}_i}\hat{\bf u} \frac{\partial}{\partial {\Psi}_j}\hat{\bf u}\right\}.
\end{split}
\end{equation}
Moreover, to simplify the notation, we use $\hat{\bf F}_{{\Psi}_i{\Psi}_j}$ to refer to the general formula of Eq. \eqref{eq:fim}. {In the following, we} calculate {both \gls{fim} 
matrices related to the \gls{sb} and \gls{db} paths.}

\vspace{2mm}
\noindent\textit{\textbf{\gls{sb} path \gls{crb}}}:
We consider that the \gls{bs} is capable of totally nullifying the echo signal component {\it c1)} and process individually the component {\it c2)} as the \gls{sb} path, as discussed previously. Hence, three variables affect this signal, as depicted before. We can evaluate the \gls{fim} {matrix} by arranging the variables contained in $\boldsymbol{\Psi}^{\rm SB}$ {applying} Eq. \eqref{eq:fim}. For illustration purposes, let's consider a single \gls{sp} scenario, where the \gls{fim} has the following form
\begin{equation}\label{eq:F_SB}
{\bf F}^{\rm SB}   = \left[
    \begin{matrix}
\hat{\bf F}_{\alpha \alpha} & \hat{\bf F}_{\alpha\tilde{\bar{\beta}}}\\
\hat{\bf F}_{\alpha\tilde{\bar{\beta}}}^\rT & \hat{\bf F}_{\tilde{\bar{\beta}}\tilde{\bar{\beta}}}
\end{matrix}\right].
\end{equation}
{We begin by computing the partial derivatives as} $\frac{\partial}{\partial \alpha}\hat{\bf u}^{\rm SB} = \bar{\beta} \dot{\bf A}(\alpha){\bf X}$ and $\frac{\partial}{\partial \tilde{\bar{\beta}}}\hat{\bf u}^{\rm SB} = [1,j]\otimes{\bf AX}$, where ${\bf A}(\alpha) = \bab(\alpha)\babt(\alpha)$, $\dot{\bf A}(\alpha) = \frac{\partial}{\partial\alpha}{\bf A}(\alpha)$.
Taking the empirical sample covariance matrix over $S$ transmitted symbols as:
\begin{equation}
{\bf R}_{\bf x} = \frac{1}{S-1} \sum_{s=1}^S {\bf x}[s]{\bf x}^{\rm H}[s], 
\end{equation}
we evaluate 
\begin{subequations}
\begin{align}
\hat{\bf F}_{\alpha \alpha} = & \frac{2S{|\bar{\beta}|}^2}{\sigma^2_n}{\rm tr}(\dot{\bf A}(\alpha){\bf R}_{\bf x}\dot{\bf A}^{\rm H}(\alpha)),\\ 
\hat{\bf F}_{\alpha\tilde{\bar{\beta}}} = & \frac{2S}{\sigma^2_n}{\rm Re} \{ {\bar{\beta}}^*{\rm tr}({\bf A}(\alpha){\bf R}_{\bf x}\dot{\bf A}^{\rm H}(\alpha))[1,j]\}, \\
\hat{\bf F}_{\tilde{\bar{\beta}}\tilde{\bar{\beta}}} =& \frac{2S}{\sigma^2_n} {\rm tr}({\bf A}(\alpha){\bf R}_{\bf x}{\bf A}^{\rm H}(\alpha)){\bf I}_2
\end{align}
\end{subequations}
Finally, we obtain the \gls{crb} of estimating angle $\alpha$ corresponds to the first element of ${{\bf F}^{\rm SB}}^{-1}$ as:
\begin{equation}\label{eq:CRBalpha}
\begin{split}\scriptsize
    [{{\bf F}^{\rm SB}}^{-1}]_{1,1} = 
    \frac{\sigma^2_n}{2S{|\bar{\beta}|}^2\bigg({\rm tr}(\dot{\bf A}{\bf R}_{\bf x} \dot{\bf A}^{\rm H})- \frac{{|{\rm tr}({\bf A}{\bf R}_{\bf x}\dot{\bf A}^{\rm H})|}^2}{{\rm tr}({\bf A}{\bf R}_{\bf x} {\bf A}^{\rm H})}\bigg)}. 
\end{split}   
\end{equation}

\vspace{2mm}
\noindent\textit{\textbf{\gls{db} path \gls{crb}}}: analogously, we can calculate the \gls{crb} for the \gls{db} {paths} by arranging the \gls{fim} with the elements contained in the set $\boldsymbol{\Psi}^{\rm DB}$ and processing {with} the re-arranged signal $\hat{\bf y}^{\rm DB}$. We can straightforwardly infer the following matrix for {single \gls{sp} target} \gls{crb} calculation\footnote{{The expression in \eqref{eq:F_DB} is valid for only one SP. Hence, it could not be apparent for the $r$ subscript since if there is more than one SP, the equations in \eqref{eq:F_DB} are not valid for all angles.}}:
\begin{equation}\label{eq:F_DB}
{\bf F}^{\rm DB} = \left[
\begin{matrix}
\hat{\bf F}_{\xi \xi} & \hat{\bf F}_{\xi\tilde{\bar{\bar{\beta}}}}\\
\hat{\bf F}_{\xi\tilde{\bar{\bar{\beta}}}}^\rT & \hat{\bf F}_{\tilde{\bar{\bar{\beta}}}\tilde{\bar{\bar{\beta}}}}
\end{matrix}\right],
\end{equation}
where 
$$
\frac{\partial}{\partial \xi} \hat{\bf u}^{\rm DB}_m = \bar{\bar{\beta}}{\rm vec}\big(\dot{\eta}_m(\xi,0){\bf A}(\alpha,0)+ \dot{\eta}_m(0,\xi){\bf A}(0,\alpha)\big){\bf X}, 
$$ and 
$$
\frac{\partial}{\partial \bar{\bar{\beta}}} \hat{\bf u} = [1,j]\otimes{\rm vec}(\eta_m{\bf A}(\alpha,0) + \eta_m{\bf A}(0,\alpha)){\bf X}
$$ for {the frequency-domain far-field pattern of the $m$-th harmonic $\eta_m$, described in Eq. \eqref{eq:STCMFm};} $\eta_m(0,\xi)$ and $\eta_m(\xi,0)$ and its derivatives assume equal complex value in both phase and magnitude due to specular reflection by the \gls{stcm}. In this context, ${\bf A}(\theta_i,\theta_j) = \bab(\phi_i)\babt(\phi_j) ={\bf A}^\rT(\phi_j,\phi_i) \triangleq {\bf A} \ \forall \ i,j$.

We can simplify the above equations to obtain $\frac{\partial}{\partial \xi} \hat{\bf u}^{\rm DB}_m = \bar{\bar{\beta}}{\rm vec}\big(\dot{\eta}_m ({\bf A} + {\bf A}^\rT){\bf X}\big) = \bar{\bar{\beta}}{\rm vec}\big(\dot{\eta}_m {\bf B}{\bf X}\big)$ and $\frac{\partial}{\partial \bar{\bar{\beta}}}\hat{\bf u}^{\rm DB}_m  = [1,j]\otimes{\rm vec}(\eta_m{\bf BX})$. Hence, supporting in Eq. \eqref{eq:fim} we can evaluate a closed-form expression as in the \gls{sb} case. Consider $\boldsymbol{\eta} = [\eta_{-mf}, \eta_{-(mf-1)}, \dots, \eta_{mf}] \in \mathbb{C}^{|\mathcal{M}|\times 1}$ and $\dot{\boldsymbol{\eta}} = [\dot{\eta}_{-mf}, \dot{\eta}_{-(mf-1)}, \dots, \dot{\eta}_{mf}] \in \mathbb{C}^{|\mathcal{M}|\times 1}$ and let:
\begin{subequations}
\begin{align}
\hat{\bf F}_{\xi \xi} & = \frac{2S{|\bar{\bar{\beta}}|}^2}{\sigma^2_n}(\dot{\boldsymbol{\eta}}^{\rm H}\dot{\boldsymbol{\eta}}){\rm tr}({\bf B}{\bf R}_{\bf x}{\bf B}^{\rm H}), \\
\hat{\bf F}_{\xi\tilde{\bar{\bar{\beta}}}} & = \frac{2S}{\sigma^2_n}{\rm Re} \{(\dot{\boldsymbol{\eta}}^{\rm H}{\boldsymbol{\eta}}) {{\bar{\bar{\beta}}}}^*{\rm tr}({\bf B}{\bf R}_{\bf x}{\bf B}^{\rm H})[1,j]\}, \\
\hat{\bf F}_{\tilde{\bar{\bar{\beta}}}\tilde{\bar{\bar{\beta}}}} & = \frac{2S}{\sigma^2_n}({\boldsymbol{\eta}}^{\rm H}{\boldsymbol{\eta}}){\rm tr}({\bf B}{\bf R}_{\bf x}{\bf B}^{\rm H}){\bf I}_2.
\end{align}
\end{subequations}
Then, the closed-form expression $[{{\bf F}^{\rm DB}}^{-1}]_{1,1}$ is {defined as:}
\begin{equation}\label{eq:CRBxi}
\begin{split}
[{{\bf F}^{\rm DB}}^{-1}]_{1,1} =  \frac{\sigma^2_n}{2S{|\bar{\bar{\beta}}|}^2{\rm tr}({\bf B}{\bf R}_{\bf x}{\bf B}^{\rm H})-\bigg(\dot{\boldsymbol{\eta}}^{\rm H}\dot{\boldsymbol{\eta}} - \frac{(\dot{\boldsymbol{\eta}}^{\rm H}{\boldsymbol{\eta}})^2}{{\boldsymbol{\eta}}^{\rm H}{\boldsymbol{\eta}}}\bigg)}.
\end{split}   
\end{equation}
{It is straightforward to} infer that the more harmonics being employed the higher the sensing capabilities, presenting a trade-off between complexity and performance. The parameter of design then becomes both $a_{pq}^m$, Eq. \eqref{eq:aFScoeff}, and the transmit signals covariance matrix ${\bf R}_{\bf x}$. Optimizing the {localization} capabilities of the system means minimizing the closed-form {expressions in} eqs. \eqref{eq:CRBalpha} and \eqref{eq:CRBxi}{, respectively.}

\vspace{2mm}
\subsubsection{{CRB on Localization}}\label{sec:Localization}
To derive the \gls{crb} on the positions of the \glspl{sp} in Euclidean coordinates we begin by computing the equivalent Fisher information of the angle parameters. For the \gls{sb} signal
\begin{equation}
    \bf{F}^{\rm EFIM}_{\alpha\alpha} = \hat{\bf{F}}_{\alpha\alpha} - \hat{\bf{F}}_{\alpha\Tilde{\bar{\beta}}} \hat{\bf{F}}^{-1}_{\Tilde{\bar{\beta}}\Tilde{\bar{\beta}}} \hat{\bf{F}}_{\alpha\Tilde{\bar{\beta}}}^\top,
\end{equation}
and for the double bounce signals
\begin{equation}
    \bf{F}^{\rm EFIM}_{\xi\xi} = \hat{\bf{F}}_{\xi\xi} - \hat{\bf{F}}_{\xi\Tilde{\bar{\bar{\beta}}}} \hat{\bf{F}}^{-1}_{\Tilde{\bar{\bar{\beta}}}\Tilde{\bar{\bar{\beta}}}} \hat{\bf{F}}_{\xi\Tilde{\bar{\bar{\beta}}}}^\rT.
\end{equation}
Then, assuming independence between the estimation of the $\alpha$ and $\xi$ angles, we define the Fisher information matrix for the position-relevant information as
\begin{equation}
    \bf{F}^{\rm EFIM} =
    \begin{pmatrix}
        \bf{F}^{\rm EFIM}_{\alpha\alpha} & 0\\
        0 & \bf{F}^{\rm EFIM}_{\xi\xi}
    \end{pmatrix}.
\end{equation}
We define the Jacobian of the mapping from the angles to the position ${\bf q}=(x,{0},z)$ as
\begin{equation}\label{eq:locCoord}
    \bf{T} =
    \begin{pmatrix}
        \frac{\partial \alpha}{\partial x} & \frac{\partial \alpha}{\partial z} \\
        \frac{\partial \xi}{\partial x} & \frac{\partial \xi}{\partial z}
    \end{pmatrix}.
\end{equation}
One can rewrite the angles as a 
the function of their position in local coordinates{, Fig. \ref{fig:bouncing}, \ref{fig:toyExample}, as:} 
$$
\alpha = \arctan \frac{q_{r,x}-q_{{\rm B},x}}{q_{r,z}-q_{{\rm B},z}}\qquad  \text{and} \qquad \xi = \arctan\frac{q_{r,x}-q_{{\rm S},x}}{q_{r,z}-q_{{\rm S},z}}
$$
Let {coordinates} $q_x$ and $q_z$ denote the point ${\bf q}(x, 0, z) ={(q_x,0,q_z)}$ in local coordinates, we can find the elements in Eq. \eqref{eq:locCoord} as
\begin{subequations}
\begin{align}
\frac{\partial \alpha}{\partial x} = & \frac{q_{{\rm B},z}-q_{z}}{(q_x-q_{{\rm B},x})^2+(q_z-q_{{\rm B},z})^2},\\
\frac{\partial \alpha}{\partial z} = & \frac{q_x-q_{{\rm B},x}}{(q_x-q_{{\rm B},x})^2+(q_z-q_{{\rm B},z})^2},\\
\frac{\partial \xi}{\partial x} = & \frac{q_{{\rm S},z}-q_z}{(q_x-q_{{\rm S},x})^2+(q_z-q_{{\rm S},z})^2}, \\ 
\frac{\partial \xi}{\partial z} = & \frac{q_x-q_{{\rm S},x}}{(q_x-q_{{\rm S},x})^2+(q_z-q_{{\rm S},z})^2}. 
    \end{align}
\end{subequations}

Then the single \gls{sp} Fisher information on the position ${\bf q}={(q_x, 0, q_z)}$ is
\begin{equation}
\bf{F}(\bf{q}) = {\bf T}^{\rT} {{\bf F}^{\rm EFIM}} {\bf T} 
\end{equation}
and the \gls{crb} on the position is ${\bf F}^{-1}({\bf q})$. We define the {\it position error bound} (PEB) as \cite{AbuShaban2018Bounds}:
\begin{equation}
{\rm PEB}({\bf q}) = \sqrt{{\rm Tr}({\bf F}^{-1}({\bf q}))}.
\end{equation}

\vspace{2mm}
\noindent{\textit{\textbf{Remark}}}: Multiple \gls{sp} \gls{peb} can be calculated via software by expanding the matrix ${\bf F}^{\rm EFIM}$ to multiple values of $\alpha$ and $\xi$, but obtaining closed form solutions is not practical in multiple SP scenarios.

\subsection{Detection and Classification}\label{sec:IV}%
{The detection problem relates not to estimating the localization of the \gls{sp} but rather to detecting its presence. This can be formulated as a hypothesis testing problem where the null hypothesis is the presence of no target and the alternate hypothesis is the presence of a target. {The }task of classification 
relates to determining which kind of {\gls{sp}} a detected target is 
{and }can also be formulated as a hypothesis-testing problem. In the following, we present the signal model in a general setting which can straightforwardly be used for both \gls{sb} and \gls{db} signals (see {eqs.} \eqref{eq:sb_signal} and \eqref{eq:db_signal}) and construct the competing hypotheses.}

\noindent The received {echo} signal from an {\gls{sp}} at position ${\bf q}_r$ under hypothesis $\mathcal{H}_i$ can be expressed as\footnote{{Notice that in \eqref{eq:Y-Hi} we assume $\lambda$ equal for every path since $f_0 << f_c$, \textit{i.e.} dropping the $m$ superscript but still the path gain is dependent on the carrier wavelength.}}
\begin{equation}\label{eq:Y-Hi}
    {\bf Y} = \beta_i {\bf H}({\bf q}_r) + \bf{N}
\end{equation}
where $ {\bf N}\sim \mathcal{CN}(0, \sigma^2_{n}{\bf I}_M)$ is \gls{awgn}, {$M$ is the dimension of the received signal}, and the path gain 
\begin{equation}
    \beta_i=G(d)\sigma_i\nu
\end{equation}
for $i=0,1,2$, where $\nu \sim \mathcal{CN}(0, \sigma^2_{\nu})$ is the fading coefficient, and $\sigma_i$ is the square root of the {radar cross section} (\gls{rcs}) with
$\sigma_0 = 0$, \textit{i.e.} absence of \gls{sp}, $\sigma_1$ refers to a NUE or UE SP type, and $\sigma_2$ refers to an object with polished surface (Obj) SP type, with RCS $\sigma_1 < \sigma_2$. 
Then, the amplitude $|\nu|$ is Rayleigh distributed with scale parameter $\sigma_{\nu}$. We know 
that $|\beta_i|$ follows a Rayleigh distribution {$R(\cdot; \varsigma)$ with scale parameter $\varsigma(\cdot) > 0$}:
\begin{equation}
{\varsigma}(\sigma_i) = G(d) \sigma_i \sigma_\nu {\left(\frac{\pi}{2}\right)^{-\frac{1}{2}}}. 
\end{equation}

\vspace{2mm}
\subsubsection{{Detection Analysis}}
{Herein, }we will assume that the position $\bs{q}{_r}$ is known, which for detection can make practical sense: we sweep through each location $\bs{q}{_r}$ and make the detection decision.  {Under such an assumption,} the maximum likelihood estimator of the channel coefficient {can be written as}:
\begin{equation}
\hat{\beta}({\bf{Y}}) = \frac{{\bf H}({\bf q}{_r})^H\bf{{Y}}}{\Vert {\bf H}({\bf q}{_r})\Vert^2}
\end{equation}
with distribution $\hat{\beta} \sim \mathcal{CN}\big(\beta_i, \, \frac{\sigma^2_{{n}
}}{\Vert \bs{H}(\bs{q{_r}})\Vert^2}\big)$. Hence, the statistic $\Tilde{\gamma} \triangleq \frac{2\Vert \bs{H}(\bs{q{_r}})\Vert^2}{\sigma^2_{{n}
}}|\hat{\beta}|^2$ follows a non-central $\chi^2$ distribution with $2$ degrees of freedom and non-centrality parameter $\mu_{\Tilde{\gamma}} \triangleq \frac{2\Vert \bs{H}(\bs{q{_r}})\Vert^2}{\sigma^2_{{n}
}}|\beta_i|^2$. Given $\beta_i$ and a detection threshold $\gamma_{\text{th}}$, the conditional detection probability is given by
\begin{equation}
p_{_{\text{D}}}(\beta_i) = f_{\Tilde{\gamma}}(\Tilde{\gamma} > \gamma_{\text{th}} | \beta_i) = Q_1\big(\sqrt{\mu_{\Tilde{\gamma}}}, \sqrt{\gamma_{\text{th}}}\big)
\end{equation}
where $Q_1(\cdot, \cdot)$ is the Marcum Q-function. The false alarm probability is related to the {detection} threshold {$\gamma_{\text{th}}$} by
\begin{equation}
    p_{_{\text{FA}}} = f_{\Tilde{\gamma}}(\Tilde{\gamma} > \gamma_{\text{th}} | \beta_i=0) = e^{-\frac{\gamma_{\text{th}}}{2}}
\end{equation}
such that $\gamma_{\text{th}} = -2\, {\ln} \, p_{_{\text{FA}}}$ for a specified false alarm probability \cite{Wymeersch2020:Detection}. In the above {expression}, we condition on the realization of the fading $\beta_i$; however, we may also be interested in the {\it marginal distribution} which can be expressed as \cite{Sofotasios2014:Marcum}
\begin{equation}\label{eq:mdPD}
p_{_{\text{D}}} = \int f_{\Tilde{\gamma}}(\Tilde{\gamma} > \gamma_{\text{th}} | \beta_i) f_{\beta_i}(\beta_i) d\beta_i.
\end{equation}

Solving the integral and simplifying the terms, 
\eqref{eq:mdPD} can be expressed as:
\begin{equation}\label{eq:mdPDeval}
p_{\textsc{d}} = \exp \left(-\frac{\gamma_{\rm th}   \sigma_n^2}{4{\Vert {\bf H}({\bf q})\Vert^2\ {\varsigma}(\sigma_i )^2+2 \sigma_n^2}}\right).
\end{equation}
Hence, the probability of detection of an \gls{sp} is dependent on the probability of a false alarm, which can be set or experimentally approximated, the noise level, the \gls{bs}-\gls{sp} array \gls{rv} of the echo signal, including the transmitted power and symbols. Finally, the \gls{sp}-\gls{bs} distance and the
\gls{rcs}, both contained in ${\varsigma}(\sigma_i )$, determine whether the \gls{bs} is capable of detecting a {scatter} target.

\subsection{{Classification Modelling in {STCM} Sensing}}

Given an observation ${\bf {Y}}(\beta_i)$, we derive the posterior probabilities for the three hypotheses. Using Bayes' formula, we have
\begin{equation}
\text{Pr}(\mathcal{H}_i|{{\bf Y}}(\beta_i)) = \frac{\text{Pr}({{\bf Y}}(\beta_i)|\mathcal{H}_i)\text{Pr}(\mathcal{H}_i)}{\displaystyle\sum_{i=0}^2\text{Pr}({{\bf Y}}(\beta_i)|\mathcal{H}_i)\text{Pr}(\mathcal{H}_i)}.
\end{equation}%
Without prior knowledge of the hypotheses, we have $\text{Pr}(\mathcal{H}_0)=\text{Pr}(\mathcal{H}_1)=\text{Pr}(\mathcal{H}_2)=\frac{1}{3}$ {in the detection/classification scenario of Fig. \ref{fig:system}, where the hypotheses $\mathcal{H}_1$ and $\mathcal{H}_2$ refer to the type of \gls{sp}, being $1$ the \gls{nue}, and $2$ the Obj; besides, hypothesis $\mathcal{H}_0$ refers to high instantaneous noise level but there are no \gls{sp} at the position analyzed, indicating a false alarm configuration. Notice that} these prior probabilities can be tuneable parameters chosen depending on the application purposes. Also, given that we have detected a point, we may assume that $\text{Pr}(\mathcal{H}_0)$ is low.

We assume that given the data ${{\bf Y}}(\beta_i)$, an estimate of $\beta_i$ is computed as $\hat{\beta}=\hat{\beta}({{\bf Y}}(\beta_i))$. Moreover, we say that $\hat{\beta} \sim f_{\hat{\beta}}(\cdot;\beta_i)$ follows a distribution with some density depending on the observation and thereby the realization of the channel $\beta_i$. The difference here compared to the setup in the detection analysis, is that we no longer assume that the position of the \gls{sp} is known, and then this uncertainty on the position will also affect the uncertainty in the estimator $\beta_i$.

To simplify the analysis, we assume that the estimator $\hat{\beta}$ is a sufficient statistic of the data $
{{\bf Y}}(\beta_i)$ with respect to the parameter $\beta_i$, such that the classification problem becomes
\begin{equation}
    \text{Pr}(\mathcal{H}_i|\hat{\beta}) = \frac{\text{Pr}(\hat{\beta}|\mathcal{H}_i)\text{Pr}(\mathcal{H}_i)}{\displaystyle\sum_{i=0}^2\text{Pr}(\hat{\beta}|\mathcal{H}_i)\text{Pr}(\mathcal{H}_i)}.
\end{equation}
We expand the conditional {probabilities} as
\begin{align}
    \text{Pr}(\hat{\beta}|\mathcal{H}_i) &= \int \text{Pr}(\hat{\beta}|\beta) \text{Pr}(\beta|\mathcal{H}_i) d\beta\\
    &= \int \text{Pr}(\hat{\beta}|\beta_i) \text{Pr}(\beta_i) d\beta_i\label{eq:hat_beta_conditionals}
\end{align}
{by} exploiting {the fact} that $\hat{\beta}$ and $\mathcal{H}_i$ are conditionally independent given $\beta$.
We remark then that $\text{Pr}(\hat{\beta}|\beta_i) = f_{\hat{\beta}}(\hat{\beta}; \beta_i)$ and $\text{Pr}(\beta_i) = \mathcal{CN}(\beta_i; 0, (G(d)\sigma_i\sigma_{\nu})^2)$. Using the {maximum a posteriori rule}, we would choose the hypothesis with the highest posterior probability, {\it i.e.}, we estimate the class as
\begin{equation}\label{eq:argmaxi}
    \hat{i} = \argmax_{i\in\{0,1,2\}} ~~ \text{Pr}(\mathcal{H}_i|\hat{\beta})
\end{equation}
and the chosen hypothesis is $\hat{\mathcal{H}} = \mathcal{H}_{\hat{i}}$. \cite{Kay1998:Detection}

\vspace{3mm}
\noindent{\textit{\textbf{Special Case: circular covariance}}}:
In this part, we assume for simplicity that the estimator of the channel coefficient follows a circular covariance\footnote{{Being rigorous, such a covariance, typically, will not be perfectly circular; however, for simplicity and expeditiousness of analyses, we have adopted such  "special case" simplification.}}, {\it i.e.}, $\hat{\beta}({\bf Y}(\beta_i)) \sim \mathcal{CN}(\beta_i, \sigma_{\hat{\beta}}^2)$ for some variance parameter $\sigma_{\hat{\beta}}^2$. 
In this case, following the same arguments as in the detection analysis, the statistic $\Tilde{\gamma} \triangleq \frac{2}{\sigma_{\hat{\beta}}^2} |\hat{\beta}|^2$ follows a non-central $\chi^2$ distribution with $2$ degrees of freedom and non-centrality parameter $\mu_{\Tilde{\gamma}} \triangleq \mu_{\Tilde{\gamma}}(|\beta_i|) \triangleq \frac{2}{\sigma_{\hat{\beta}}^2}|\beta_i|^2$. The density function for $\Tilde{\gamma}$ is then
\begin{equation}
    f_{\Tilde{\gamma}}(\Tilde{\gamma}; \mu_{\Tilde{\gamma}}) = \frac{1}{2} I_0\Big(\sqrt{\mu_{\Tilde{\gamma}} \Tilde{\gamma}}\Big) \exp \Big\{-\frac{\Tilde{\gamma} + \mu_{\Tilde{\gamma}}}{2}\Big\}
\end{equation}
By a variable transformation, we find the density for $|\hat{\beta}|$ as
\begin{align}
    f_{|\hat{\beta}|}(|\hat{\beta}|; \mu_{\Tilde{\gamma}}) &= \Big|\frac{d \Tilde{\gamma}}{d |\hat{\beta}|}\Big| f_{\Tilde{\gamma}}\Big(\frac{2|\hat{\beta}|^2}{\sigma_{\hat{\beta}}^2}; \mu_{\Tilde{\gamma}}\Big) \nonumber \\
    &= \frac{2|\hat{\beta}|}{\sigma^2_{\hat{\beta}}} I_0\Bigg(\sqrt{2 \mu_{\Tilde{\gamma}}}\frac{|\hat{\beta}|}{\sigma_{\hat{\beta}}}\Bigg) \exp\Bigg\{-\Big(\frac{|\hat{\beta}|^2}{\sigma^2_{\hat{\beta}}} + \frac{\mu_{\Tilde{\gamma}}}{2}\Big)\Bigg\}.
\end{align}
where $I_0$ is the zeroth order Bessel function of the first kind.

Now, we denote by $R(\cdot; {\varsigma})$ the density of the Rayleigh distribution with scale parameter ${\varsigma(\cdot) }> 0$. Using now the statistic $|\hat{\beta}|$, we express the posterior {probability} as
\begin{equation}
\text{Pr}(\mathcal{H}_i\:|\:|\hat{\beta}|) = \frac{\text{Pr}(|\hat{\beta}||\mathcal{H}_i)\text{Pr}(\mathcal{H}_i)}{\displaystyle\sum_{i=0}^2\text{Pr}(|\hat{\beta}||\mathcal{H}_i)\text{Pr}(\mathcal{H}_i)}.
\end{equation}
We know from earlier that $|\beta|$ is Rayleigh distributed, hence for $i=0,1,2$
\begin{equation}
    \text{Pr}(|\beta||\mathcal{H}_i) = R(|\beta|;{\varsigma}(\sigma_i))
\end{equation}
where we put $\sigma_0 \triangleq 0$ and $R(|\beta|;0)$ is used to denote the Dirac delta distribution centered at $0$. This results in the following conditional probabilities
\begin{align}\label{eq:SCClass}
    \text{Pr}(|\hat{\beta}||\mathcal{H}_i) &= \int f_{|\hat{\beta}|}(|\hat{\beta}|; \mu_{\Tilde{\gamma}}(|\beta|)) R(|\beta|;{\varsigma}(\sigma_i)) d|\beta|\nonumber\\
    &= \frac{2|\hat{\beta}|}{2\varsigma(\sigma_i)^2 + \sigma_{\hat{\beta}}^2} \exp \Big(-\frac{|\hat{\beta}|^2}{2\varsigma(\sigma_i)^2+\sigma_{\hat{\beta}}^2}\Big).
\end{align}
Intuitively, we can understand that when the estimate $|\hat{\beta}|$ is higher, we will have a higher probability that the target is an Obj; however, if we at the same time have a large covariance, then the uncertainty about this labeling is high, and we have decreased probability of being an Obj. This can express a classification that not only labels the target but also provides a level of certainty about the label. \cite{Khawar2015:Detection}
%
%
\vspace{2mm}
\subsubsection{{Classification: system performance overview}}
Taking a system-level performance view on the classification, we are interested in the probability of correctly classifying the target given the true class of the target, which we write as
\begin{equation}
    \text{Pr}(\hat{\mathcal{H}} = \mathcal{H}_i | \mathcal{H}_j)
\end{equation}
for $i,j=0,1,2$. The statistic we use to base the classification decision is $\hat{\beta}$. Therefore, we expand the probability as
\begin{equation}
    \text{Pr}(\hat{\mathcal{H}} = \mathcal{H}_i | \mathcal{H}_j) = \int \text{Pr}(\hat{\mathcal{H}} = \mathcal{H}_i | \hat{\beta}) \text{Pr}(\hat{\beta}|\mathcal{H}_j) d\hat{\beta}.
\end{equation}
We know the probability $\text{Pr}(\hat{\beta}|\mathcal{H}_j)$ from the {conditional probabilities} expansion, Eq.~\eqref{eq:hat_beta_conditionals}. On the other hand, the event $\hat{\mathcal{H}} = \mathcal{H}_i | \hat{\beta}$ is deterministic in nature, as the classification decision given the statistic $\hat{\beta}$ is deterministic. Specifically,
\begin{equation}
    \text{Pr}(\hat{\mathcal{H}} = \mathcal{H}_i | \hat{\beta}) =
    \begin{cases}
        1 & \text{if} ~~ \frac{\text{Pr}(\hat{\beta}|\mathcal{H}_i)\text{Pr}(\mathcal{H}_i)}{\text{Pr}(\hat{\beta}|\mathcal{H}_j)\text{Pr}(\mathcal{H}_j)} > 1, ~\forall j\neq i,\\
        0 & \text{otherwise}.
    \end{cases}
\end{equation}

\vspace{2mm}
\subsubsection{{Classification: information fusion}}
Assume that we estimate the channel coefficient for both non-{\gls{stcm}} 
and through the {\gls{stcm}}, resulting in two estimators $\hat{\beta}^{{\textsc{r}}}$, which is an estimate of $\bar{\bar{\beta}}(\sigma_i)$ and $\hat{\beta}^{{\textsc{n}}}$ which is an estimate of $\bar{\beta}(\sigma_i)$. Then, the posterior {probability} for classification is $\text{Pr}(\mathcal{H}_i | \hat{\beta}^R, \hat{\beta}^N)$ and the likelihood is
\begin{equation}
{\text{Pr}}(\hat{\beta}^{{\textsc{r}}}, \hat{\beta}^{{\textsc{n}}} | \mathcal{H}_i) = {\text{Pr}}(\hat{\beta}^{{\textsc{r}}} | \mathcal{H}_i) \, {\cdot}\, {\text{Pr}}(\hat{\beta}^{{\textsc{n}}} | \mathcal{H}_i)
\end{equation}
assuming independence of the small-scale fading and assuming known data association, i.e., knowing that $\hat{\beta}^{{\textsc{r}}}$ and $\hat{\beta}^{{\textsc{n}}}$ are estimates of the channel coefficient for the same target.

\section{{Numerical Illustration}}\label{sec:V}
Throughout this section, we carry out simulations under the parameters of Table \ref{tab:SimulationParameters} otherwise specified. We employ a DFT-based product {to represent the} beamforming vector ${\bf w}$ and signals ${\bf s}$ defined as:
\begin{equation}
    \mathbf{X} = {\bf w s} = \{\boldsymbol{\mu}_i\}^{\sqrt{M}}_{i=1} \otimes \{\boldsymbol{\nu}_{j}\}^{\sqrt{M}}_{j=1},
\end{equation}
where
\begin{align}
    \boldsymbol{\mu}_i = \left[1, e^{\frac{2\pi(i-1)}{\sqrt{M}}}, \dots , e^{\frac{2\pi(i-1)(\sqrt{M}-1)}{\sqrt{M}}}\right],\\
    \boldsymbol{\nu}_j = \left[1, e^{\frac{2\pi(j-1)}{\sqrt{M}}}, \dots , e^{\frac{2\pi(j-1)(\sqrt{M}-1)}{\sqrt{M}}}\right],
\end{align}
{ where in this representation, the symbols dimension $S=M$,}  
guaranteeing that the pilot signals are orthogonal, thus having their covariance matrix approximately a weighted identity, which is known for having higher performance than non-orthogonal signals \cite{Bekkerman2006}. We analyze the {system} feasibility two-fold, {by evaluating the system localization and the detection/classification capabilities under single and multiple target scenarios.}

\begin{table}[!tbhp]
\caption{Simulation Parameters}
\centering
\begin{tabular}{l|l}\hline
 \bf Parameter      & \bf Value  \\ \hline
{Number of scatter points (SP)} & $|\mathcal{R}| \in \{1,2, 10\}$\\
Number of UEs & $K=1$\\
\hline
 \gls{bs} antennas  & $M=16$     \\
 \gls{stcm} elements & $N = 64$  \\
 Analyzed harmonics & $m_f = \{3,4,5\}$\\
 {Coding 
 length} & $L = 8$ \\
 Coding period      & $T_0 = 2 \ \mu$s \\
 Coding {rate} 
 & {$f_{0} = 500$} kHz \\
 Carrier frequency  & $f_c = 10$ GHz \\
 Path loss exponent & $\iota^2 = 4$ \\
Total power across $S$ symbols  & $\|{\bf X}\| = 12$ dBm \\
 Noise power        & $\sigma^2_n = -120$ dBm\\
 \gls{nue} \gls{rcs}&  $\sigma_r= 1$ dB $\cdot$ m$^2$\\
 Obj \gls{rcs}      &  $\sigma_r = 17$ dB$\cdot$m$^2$ \cite{kim2022}\\
 \hline
 {Physical area (Fig. 1)} &  {$160\times 100$ m$^2$}\\
 {STCM center localization} & {(0, 0, 100) m}\\
 {ULA center localization} & {(0, 0, 0) m}\\
\hline
\end{tabular}
\label{tab:SimulationParameters}
\end{table}

\noindent\textbf{\textit{Cramér-Rao Bounds}}
Suppose the \glspl{sp} have a particular characteristic of reflecting an incident wave with full power $(\sigma_r$ = 0 dB$\cdot$m$^2$, $\sigma^2_{\nu} = 0$). Calculating the \gls{crb} {for the attainable variance of the estimation of the angle} $\alpha$ and $\xi$ is enough information to retrieve an \gls{sp} localization {(Eq. \eqref{eq:angRelation}) {with} accuracy}. We start our assessment by defining $|\mathcal{M}| = 7$, \textit{i.e.} $m_f = 3$, and $L = 8$, using the {procedure} sequence {as suggested} 
in Fig. \ref{fig:system}; {we} evaluate the \gls{crb}s on \gls{stcm}-assisted \gls{isac} system for the following cases, from the simplest to more complex:
1) single target;
2) double target, and
3) multiple targets. {Moreover, in} 1), 2) and 3) {scenarios}, we use only \gls{sb} to calculate the $\alpha$ \gls{crb} {variance} and \gls{db} for $\xi$ \gls{crb} {variance}. Next, we show the attainable localization accuracy in this novel approach by evaluating the \gls{peb}, Sect. \ref{sec:Localization}, for the three scenarios.

\begin{figure}[!htbp]
\centering
\subfloat[Cramer Rao Bounds w.r.t. $\alpha$.]{\includegraphics[width=.495\linewidth]{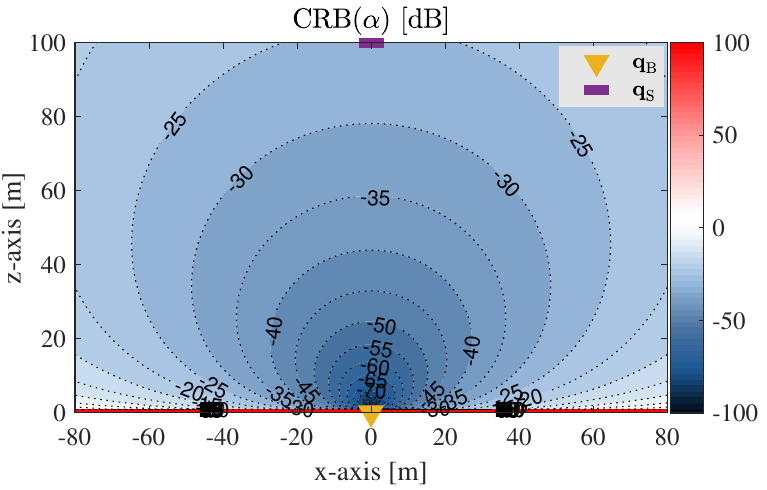}\label{fig:CRBalphaST}}\hfill
\subfloat[Cramer Rao Bounds w.r.t. $\xi$.]{\includegraphics[width=.495\linewidth]{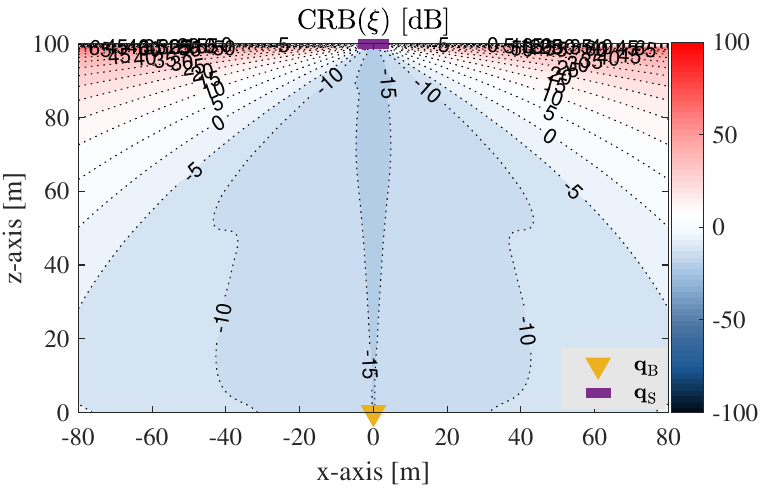}\label{fig:CRBxiST}}
\caption{\small Single target estimation Cramér Rao Bounds. The contour plot, scaled in dB, shows the variance of the estimator given the location of an \gls{sp} in ${\bf q}_r = (x, 0, z), x\in[-80:80], z\in[0,100]$ meters.} 
\label{fig:CRBST}
\end{figure}

\vspace{2mm}
\noindent\textit{\textbf{Single target}}: on a single target, it is expected the highest performance in estimating {angle} $\alpha$, as depicted in {Fig. \ref{fig:toyExample}, mainly when the \gls{sp} is close the \gls{bs}} and {under} the lowest at broadside angles. {Besides, for estimating of $\xi$ angle}, the behavior of the harmonics is the key indicator. We set $m_f = 3$ and both {\glspl{crb}} calculations are depicted in Fig. \ref{fig:CRBST}.
This result {reveals} reliable capacity {of the proposed \gls{stcm}-based method} in estimating a single \gls{sp} position through retrieving intrinsic signal properties and its synthetic {harmonic} frequencies generated by {switching the impinged signal at the} metasurface.

The result of Fig. \ref{fig:CRBalphaST} reveals that the {\gls{crb} variance for the $\alpha$ angle estimation} decreases as $d_r$ increases, except for extreme broadside angles. Hence, the {\it localization} performance is bounded {by} the \gls{snr} of the signal, involving the propagation characteristics, signals, and transmit power. {Moreover, the \gls{crb} for the $\xi$ angle estimation} is shown to be directly bounded {by} the directivity of the generated harmonics {in the  \gls{stcm}}. One can {infer} that when 
{$\xi \approx 0$} degrees, {the variance of the $\xi$ estimates achieves its lowest value}. {At the same time, the estimator's variance} is higher than 1 when $\xi \gtrapprox 60$ degrees, due to the reflected signal strength in those locations being very attenuated with the employed switching pattern and the number of harmonics \cite{Zhang2018}{. Hence, those parameters associated with switching patterns and the number of harmonics as described in eqs. \eqref{eq:STCMFm} and \eqref{eq:aFScoeff} can be optimized via Eq. \eqref{eq:CRBxi}.} 

\begin{figure}[!htbp]
    \centering
    \includegraphics[width = 1\linewidth]{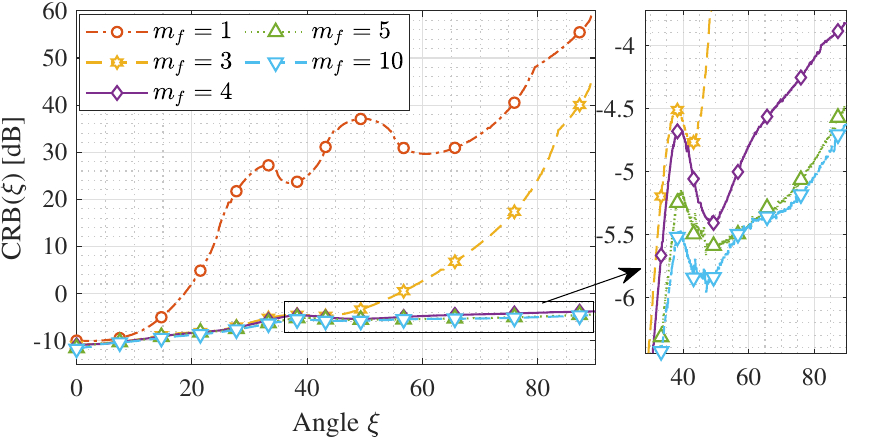}
    \caption{Highest order harmonic $m_f$ \textit{versus} \gls{crb}($\xi$) in [dB], eq. \eqref{eq:CRBxi}, as a function of the \gls{stcm}-\gls{sp} angle $\xi$.}
    \label{fig:CRBxiSTmfComp}
\end{figure}

We stick to the number of harmonics. In Fig. \ref{fig:CRBxiSTmfComp}, we analyze the CRB variance of $\xi$ using more frequency harmonics of higher orders. 
%
Notice that including more harmonics above $m_f =3$ can increase the localization {accuracy} to the whole analyzed environment (\gls{crb}$(\xi)<1, \ \forall \xi$), but does not increase significantly the result for angles lower than $50^{\rm o}$. Moreover, we can see that using more than four harmonics, $m_f>4$, does not increase significantly the performance, with just a slight variation of $1$ dB when comparing $m_f =4$ to $m_f=5$ \gls{crb} curves and much lower {improvement} for $m_f = 10$ in comparison to $m_f = 5$.  Thus, under this switching pattern, the efficient configuration in complexity and performance is attained when $m_f = 4$ positive and negative harmonics are deployed. Next, we stress the capability of estimating double and multiple target positions.

\begin{figure}[htbp!]
\centering
\subfloat[Cramer Rao Bounds w.r.t. $\alpha$ for $|\mathcal{R}|=2$.]{\includegraphics[trim = 2mm 2mm 0mm 7mm,clip,width=1\linewidth]{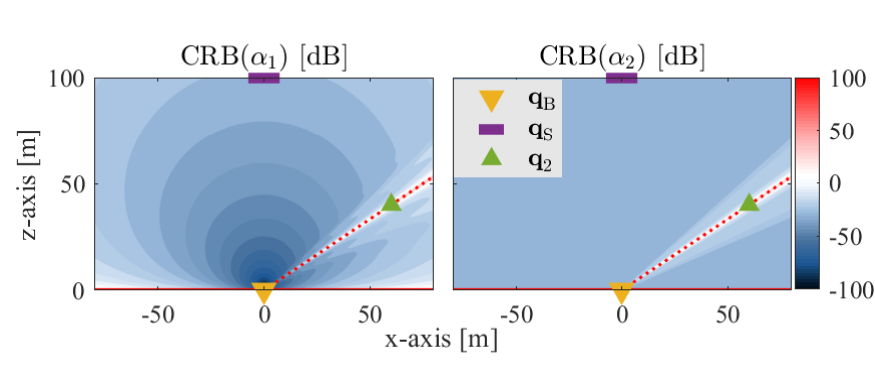}\label{fig:CRBalphaDT}}\vfill
\subfloat[Cramer Rao Bounds w.r.t. $\xi$ for $|\mathcal{R}|=2$.]{\includegraphics[trim = 2mm 2mm 0mm 7mm,clip,width=1\linewidth]{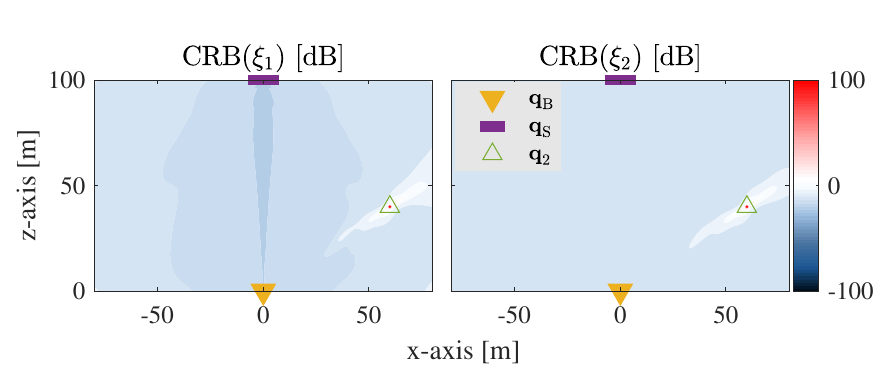}\label{fig:CRBxiDT}}
\caption{\small Double target estimation Cramér Rao Bounds. The contour plot shows the variance of the estimator for \gls{sp}$_1$ (left plots) in ${\bf q}_1 = [x,0,z], x\in[-80,80], z\in[0,100]$ given the location of an \gls{sp}$_2$ in ${\bf q}_2 = [60, 0, 40]$ meters. Right plots is the \gls{crb}s for \gls{sp}$_2$.}
\label{fig:CRBDT}
\end{figure}

\vspace{2mm}
\noindent\textbf{\textit{Double target}}: 
we want to formulate the \gls{crb} to evaluate the effect of the mutual interference and limitations of localization estimation in the proposed {\gls{stcm}-assisted \gls{m-mimo}} system. In this setup, we fix an \gls{sp}$_2$ location in ${\bf q}_2 = [60,0,40]$ and calculate the \gls{crb} for \gls{sp}$_1$ in ${\bf q}_1 = [q_x,0,q_z]$ position and the \gls{crb} for \gls{sp}$_2$ when \gls{sp}$_1$ is in position ${\bf q}_1 = [q_x,0,q_z]$, respectively. We do this calculation {numerically by} expanding the \gls{fim}s, Eq. \eqref{eq:F_SB} and Eq. \eqref{eq:F_DB} {for the \gls{sb} and \gls{db} configurations, respectively}. This result is presented in Fig. \ref{fig:CRBDT}.

Fig. \ref{fig:CRBDT} reveals that the angles estimation \gls{crb} when $|\mathcal{R}|> 1$ are bounded to the spatial distribution of them. The mutual interference is higher when the \glspl{sp} share the same \gls{bs}-\gls{sp} angle, $\alpha$, and the estimation of $\alpha_r, \ r =\{1,2\}$, {indicates to be either unfeasible or} very imprecise, Fig. \ref{fig:CRBalphaDT}. The estimation of $\xi_r$ is bounded by a region when one \gls{sp} is placed next to the other, Fig. \ref{fig:CRBxiDT}. However, to precisely localize the \glspl{sp}, the estimator's variance must be low for both parameters, showing both circumstances bound the performance.

\vspace{2mm}
\noindent{\textbf{\textit{Multiple targets:}} let's stress the system capabilities with a massive number of targets, $|\mathcal{R}|=10$, where nine targets are fixed and deployed equally spaced in a manner to cover the whole range of \gls{bs}-\gls{sp} angles, excluding the extremes broadsides, {while} evaluating the \gls{crb} {variance estimate} for a target in ${\bf q}_1 = [x,0,z]$ location. 
Fig. \ref{fig:CRBMT} reinforce physical limitation on precisely estimating the localization when more than one \gls{sp} is deployed at the {\it same angle}, Fig. \ref{fig:CRBalphaMT}, or {\it same region}, Fig. \ref{fig:CRBxiMT}. In the first case, the contour plot in  Fig \ref{fig:CRBalphaMT} reveals that the broadside \glspl{sp} are the ones that make the localization estimation inaccurate or even infeasible (white-to-red and red areas), affecting a much larger region than the rest. In the case of $\xi$ covariance estimation, Fig. \ref{fig:CRBxiMT} {corroborates} the \gls{stcm} can overcome the limitations related to the angles and is bounded to a region that is larger for broadside \glspl{sp} too. Hence, one can conclude that those estimations may not be precise in such locations.  

\begin{figure}[htbp!]
\centering
\subfloat[{$|\mathcal{R}|=10$ CRB \gls{bs}-\gls{sp} angle CRB}]{\includegraphics[trim = 0mm 5mm 0mm 0mm,width=.495\linewidth]{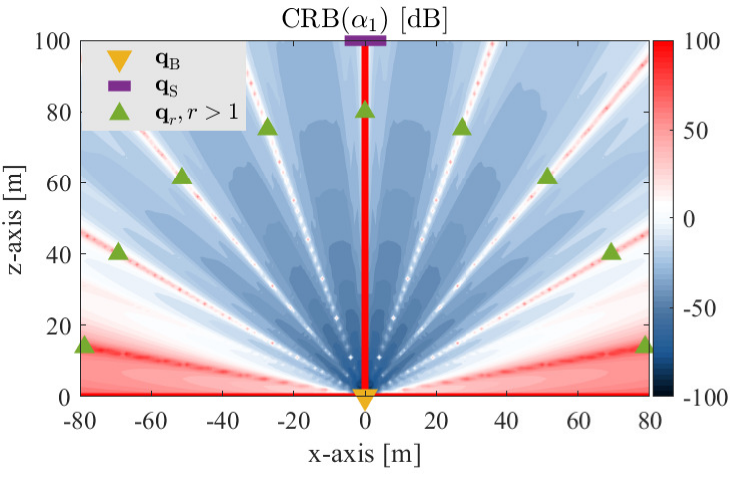}\label{fig:CRBalphaMT}}\hfill
\subfloat[{$|\mathcal{R}|=10$ \gls{stcm}-\gls{sp} angle CRB}]{\includegraphics[trim = 0mm 5mm 0mm 0mm, width=.495\linewidth]{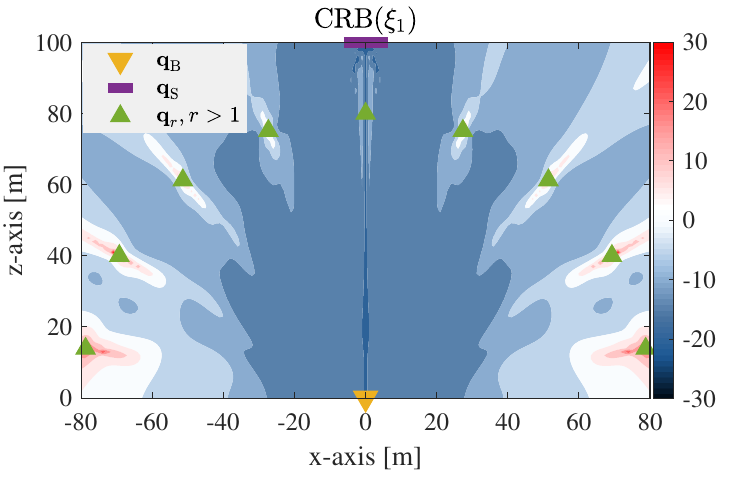}\label{fig:CRBxiMT}}
\caption{Cramer Rao Bounds for (a) $\alpha$ and (b) $\xi$ angles in multiple target $|\mathcal{R}|=10$ case.}
    \label{fig:CRBMT}
\end{figure}

To {comprehensively} evaluate the \textit{localization} capabilities {of the proposed method}, {in the following subsections we reinterpret the information of the \gls{crb} variance of the angles estimates} to local coordinates.

\subsection{Localization performance}
We can see from previous {numerical result} calculations that an optimal estimator is capable of estimating both angles studied here with a low error covariance, but it does not give us precise information on capabilities. To evaluate the local coordinates localization performance, we translate the information in angles relationship to the \gls{peb}, Sect. \ref{sec:Localization}. The \gls{peb} gives the precise bound of localization in meters considering the information contained in the channel. Hence, let us present in Fig. \ref{fig:PEBAll} the \gls{peb} for the single, Fig \ref{fig:PEB_1SP}, double, Fig \ref{fig:PEB_2SP}, and multiple targets, \ref{fig:PEB_10SP}, scenarios shown previously.

\begin{figure*}[!bhtp]
\centering
\subfloat[{\gls{peb} when a single \gls{sp} is positioned at ${\bf q}_r =(x,z), x\in[-80,80]$ m and $z\in[0,100]$ m.}]{\includegraphics[width=0.32\linewidth]{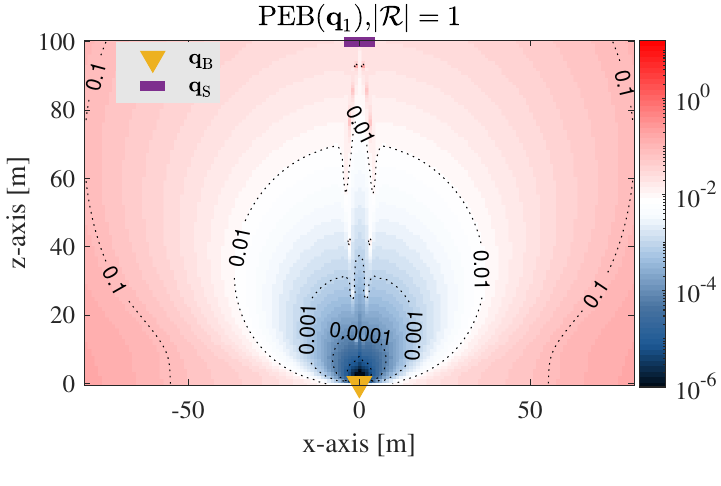}\label{fig:PEB_1SP}}\hfill
\subfloat[{\gls{peb} when an \gls{sp} is positioned at ${\bf q}_1 =(x,z), x\in[-80,80]$ m and $z\in[0,100]$ m with a fixed \gls{sp} at ${\bf q}_2 =(60,40)$ m.}]{\includegraphics[width=0.32\linewidth]{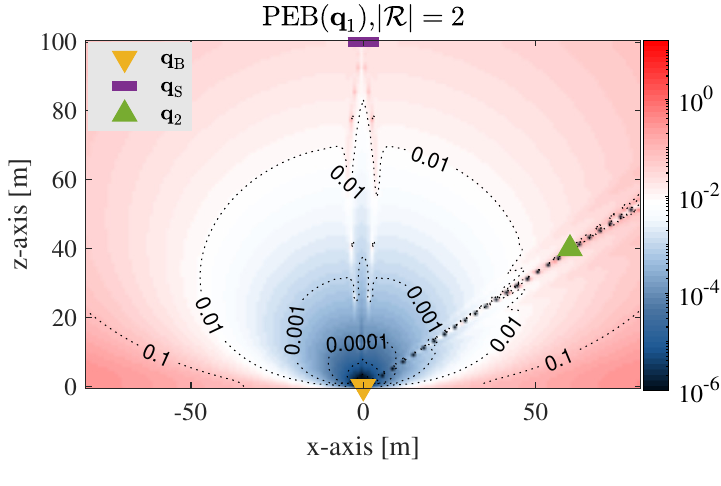}\label{fig:PEB_2SP}}\hfill
\subfloat[{\gls{peb} when an\gls{sp} is positioned at ${\bf q}_1 =(x,z), x\in[-80,80]$ m and $z\in[0,100]$ m with a set of fixed \glspl{sp} equidistributed around the \gls{bs} (see Fig. \ref{fig:CRBMT})}]{\includegraphics[width=0.32\linewidth]{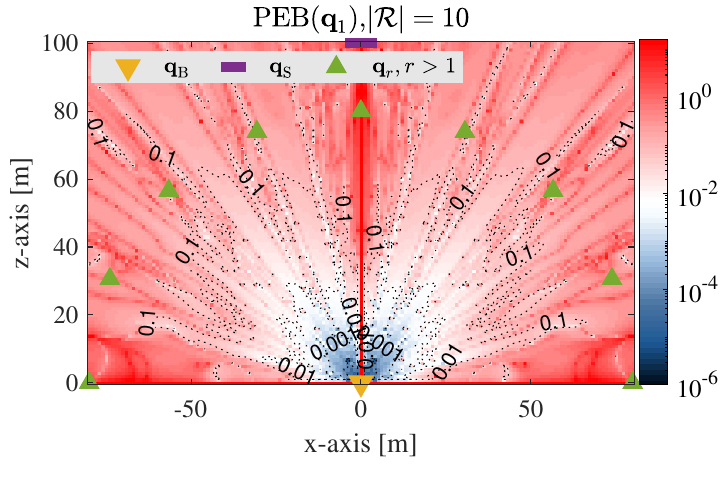}\label{fig:PEB_10SP}}
\caption{Position Error Bounds for the various number of \glspl{sp}.}
\label{fig:PEBAll}
\end{figure*} 

Fig. \ref{fig:PEBAll} {reinterprets} the \gls{crb} covariance into localization accuracy in meters where the behavior of the angles estimation variances are reflected in the \gls{sp} position estimation. The localization performance decreases with $d_r$ {increasing}, Fig. \ref{fig:PEB_1SP}, from micro centimeter level to centimeter level, with a large region below subcentimeter level and, in the worst case for a broadside of both \gls{bs} and \gls{stcm}, to tenths-of-centimeter level. With the addition of more \gls{sp}, we can see that the \gls{peb} $\gg 1$ when $\alpha_1 = \alpha_2$, Fig. \ref{fig:PEB_2SP}, being the most significant limitation of the system. Moreover, when $|\mathcal{R}| = 10$, the localization is degraded compared to previous cases but still achieves a tenth-of-centimeter \gls{peb} level or below for a big area. Hence, the presence of the \gls{sp} can be viewed as an interference, and this performance may be increased by the successive cancelation of signals or employing different combiners, e.g., zero forcing, once an \gls{sp} is localized to eliminate its interference from others \gls{sp} signal. Thus, this is a subject for future work.

\subsection{Detection and Classification performance}
From the {numerical simulations presented} in previous subsections, one can affirm that, under certain limitations, the \gls{bs} is capable of recovering not only the \gls{aoa} of the \glspl{sp} but also their position due to characteristics brought by the \gls{stcm} once they are detected. However, the detection is not immediate and is subject to false alarms. To evaluate the {detection capability}, let us rely on the framework presented in Section \ref{sec:IV}. 

\vspace{2mm} 
\noindent {\bf Detection}: recall Section \ref{sec:IV}, where we assume a known targets' location ${\bf q}$. We want to evaluate the theoretical detection probability by evaluating Eq. \eqref{eq:mdPDeval} under 
parameters 
$\sigma_i$ depicted in Table \ref{tab:SimulationParameters} resembling the \gls{nue} and Obj detectability. From Eq. \eqref{eq:mdPDeval}, it is straightforwardly inferred that the detection probability is dependent on the noise power, a threshold for detection, the channel {state} condition, including transmit and combining/receiving beamforming and pilot symbols, and the specific targets' \gls{rcs}. Let 
${\bf H}({\bf q})$ be defined as
\begin{equation}\label{eq:HqDef}
        {\bf H}({\bf q}) = {\bf Z} \bab(\angle {\bf q})\babt(\angle {\bf q}) {\bf X}
\end{equation}
\noindent with ${\bf Z}$ being a special case of combining beamformer having the same dimension as ${\bf X}$ employed in the same manner as \textit{de-spreading} operation in channel estimation \cite{Marzetta2016Book}. 
We use only the direct component \textit{c2)} since $\bar{\beta}_r \gg \bar{\bar{\beta}}_r$. We compare two configurations for de-spreading, being \textit{i)} ${\bf Z} = {\bf 1}_M$ and \textit{ii)} ${\bf Z} ={\bf X}^{\rm H}$. The position ${\bf q}$ \textit{versus} probability detection $p_D$ is depicted in Fig. \ref{fig:theoretical_pD}. 

When using the \textit{naive} combiner \textit{i)}, Fig \ref{fig:theoretical_pD}, \textit{left} subplots, the performance is greatly degraded in comparison with \textit{ii)}, \textit{right} subplots, which is consistent with physical characteristics of the system. Also, we can notice that the higher the \gls{rcs}, the higher the detection probability, which can be compared by looking at the color bar ranges depicted in the graphics.

\begin{figure}[!htbp]
\centering
\includegraphics[width = .98\linewidth]{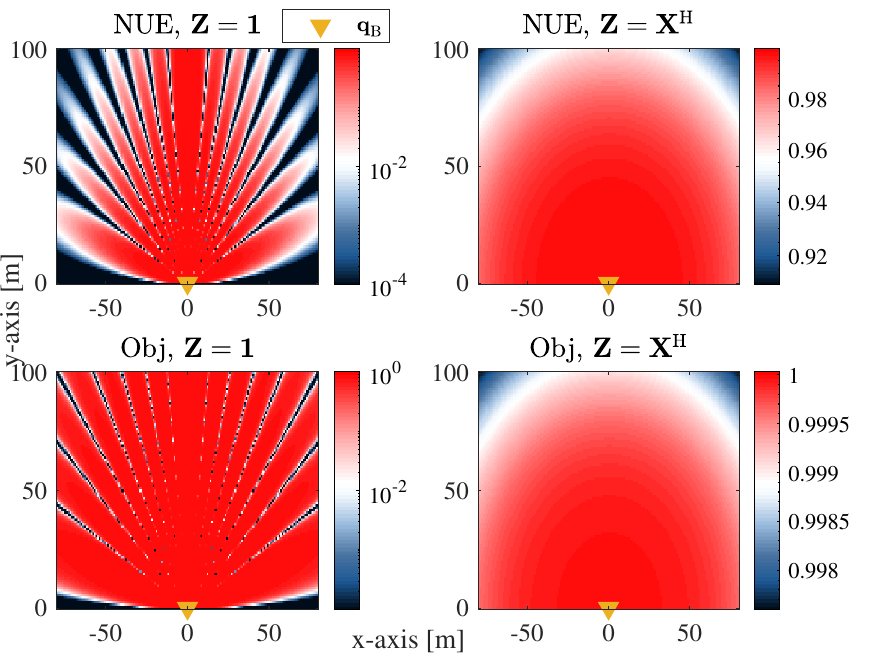}
\caption{Theoretical detection probability comparison by \gls{sp} type and combiner under $p_{\rm FA} = 10^{-4}$.}
\label{fig:theoretical_pD}
\end{figure}

\vspace{2mm}
\noindent{\bf Classification:} given an \gls{sp} were detected, we can classify whether it is a false alarm, i.e. $\hat{i}=0$, an \gls{nue}, $\hat{i} = 1$, or an Obj, $\hat{i}=2$, recalling from the nomenclature defined in Section \ref{sec:IV}. We will consider the \textit{Special Case: circular covariance} and evaluate Eq. \eqref{eq:SCClass}.

{We resort to Monte Carlo simulations to present the results in this stage.} The setup {includes 
a) for} every point in the enclosed space of ${\bf q}_x \in [-80,80], {\bf q}_y \in [0,100]$ we simulate the existence of one \gls{sp} of each type separately. Then, we evaluate Eq. \eqref{eq:SCClass} and solve Eq. \eqref{eq:argmaxi} to evaluate the classification probability of each type of \gls{sp}. The {numerical} results are depicted in Fig. \ref{fig:classAnalysis}.

\begin{figure}[!b]
\centering
\includegraphics[width = .98\linewidth]{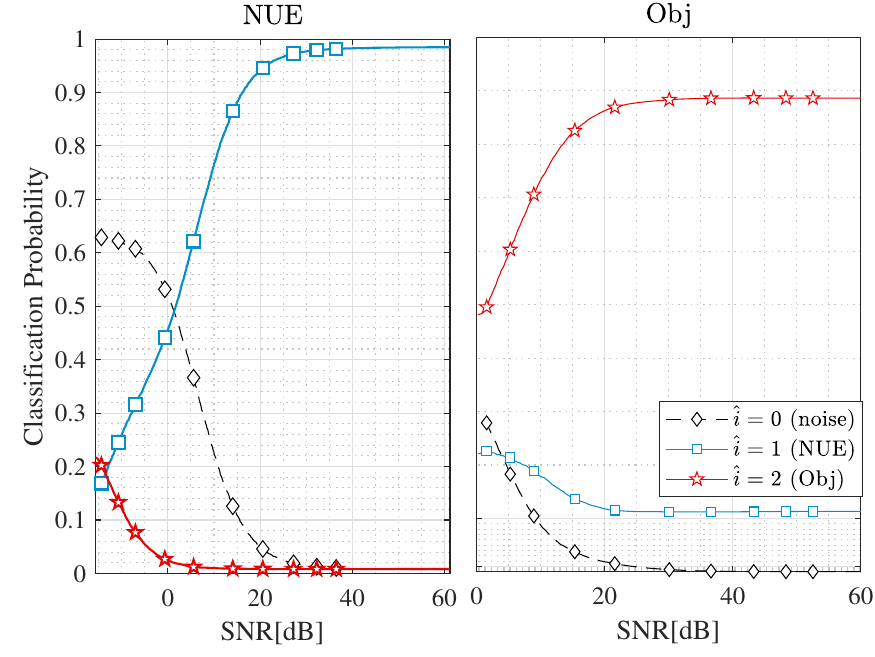}
\vspace{-1mm}
\caption{{Classification analysis for an impinging signal reflected on the two types of \gls{sp}.}}
\label{fig:classAnalysis}
\end{figure}

{Fig. \ref{fig:classAnalysis} shows the \gls{sp} mean \gls{snr} $\frac{|{\rm vec}(\bar{\beta}_r{\bf H})^{\rm H} {\rm vec}(\bar{\beta}_r{\bf H})|}{\sigma_n^2}$  \textit{versus} classification probability of a signal reflected on an \gls{nue} (left) and in an \gls{sp} (right). The simulations were performed in the $100 \times 160$ squared-meter scenario, with the curves obtained by averaging $10^4$ runs of the algorithm. With \gls{snr} higher than $20$ dB, the probability of correct classification of \gls{nue} signal is above $90$\% with a plateau of $98.4926$\% after $\approx 35$ dB, while for Obj the plateau is on $88.636$\% for \gls{snr}s higher than $30$ dB. The probability of misclassification between \glspl{sp} is low for both cases, approximating zero for high \gls{snr}s in the \gls{nue} case, but with a floor of $11.2256$\% in Obj case. This occurs in higher \gls{snr}s, \textit{i.e.} when the \gls{sp} is nearer to the \gls{bs}, being this the region where the Rayleigh distributions of both signals have higher overlap. Below zero \gls{snr} the classification as noise is most likely to occur.}

\subsection{Sensing with RIS}
{Finally,} we evaluate whether it is possible to perform the sensing framework
proposed herein by deploying {an ordinary metasurface, {\it i.e.} that does not perform space-time coding} 
{within a single transmission interval}. Hence, we {re}formulate the received signal exchanging the \gls{stcm} non-linear space-frequency scattering and substitute {by} 
the \gls{ris} reflecting framework {with a fixed elements' configuration}. Hence, Eq. \eqref{eq:echo} is adapted specifically for angles {$\alpha$} and $\xi${, maintaining the static characteristic analyzed before, thus $d_m(\tau)=1$.} 
\begin{equation}
\begin{split}
{\bf Y}^{\rm echo}_{\rm RIS}  \triangleq 
\bigg(\underbrace{\bar{\beta}_{\rS} \bab(0) (\baRt(0) \boldsymbol{\Omega}\baR(0))\babt(0)}_{\text{{\it c1)} BS-RIS-BS}}\\
+ \sum\limits_{r\in\mathcal{R}} \underbrace{\bar{\beta}_{r} \bab(\alpha)\babt(\alpha)}_{\text{{\it c2)} BS-SP-BS}}  \\
+ \sum\limits_{r\in\mathcal{R}} \underbrace{\bar{\bar{\beta}}_r\bab({\alpha}) (\baRt({\xi}) \boldsymbol{\Omega}\baR(0))\babt(0)}_{\text{{\it c3)} BS-RIS-SP-BS}}\\
+\sum\limits_{r\in\mathcal{R}} \underbrace{\bar{\bar{\beta}}_{r}\bab(0) (\baRt(0)) \boldsymbol{\Omega}\baR({\xi}))\babt ({\alpha})}_{\text{{\it c4)} BS-SP-RIS-BS}}\bigg)    
    {\bf X} + {\bf N},
\end{split}
\end{equation}
where $\baR(\cdot) \in \mathbb{C}^{N\times 1}$ denotes the \gls{ris} array \gls{rv} and $\boldsymbol{\Omega}={\rm diag}(\boldsymbol{\omega})$, where $\boldsymbol{\omega} = [\omega_1, \omega_2,\dots,\omega_N]$ is the \gls{ris} phase{-shift} profile. Straightforward, we can retrieve the \gls{crb} using the same procedure as described in Sect. \ref{sec:III}. We suppress the derivations and stick to the result presented in Fig. \ref{fig:CRBwRIS}.

\begin{figure}[!htbp]
\centering
\includegraphics[width=\linewidth]{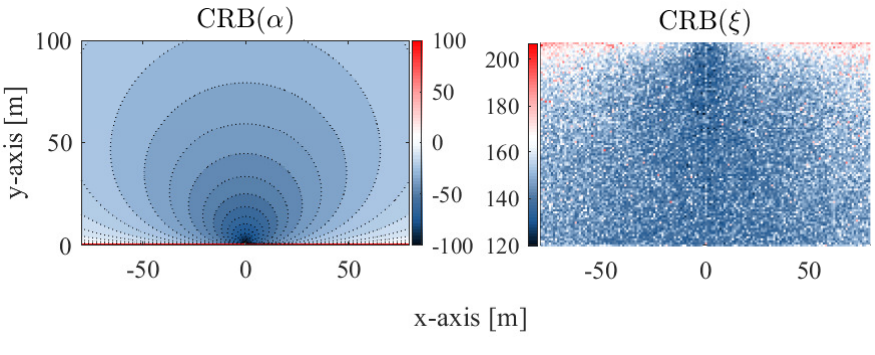}
\caption{\gls{crb} in linear \gls{ris} assisted sensing.}
\label{fig:CRBwRIS}
\end{figure}

The behavior presented in Fig. \ref{fig:CRBwRIS} shows the incapability of applying the linear metasurface in the proposed approach. The surfaces obtained indicate that it is not possible to retrieve the \gls{ris}-\gls{sp} angle since an estimator variance is as high as 120 dB in the best case. {Hence, to produce similar results as obtained with \gls{stcm}, it would need at least $|\mathcal{M}|$ configurations on the {conventional} \gls{ris} phases profile, thus a huge amount of symbols, to reproduce the frequency-space scattering {setups,} in which the sensing framework is based. Such a method of operation demands a large amount of resources in the control channel and timing, and may not be feasible to be implemented {directly in the ISAC systems} due to physical constraints on the channels' coefficients during a frame.} The detection and classification do not rely on harmonic frequencies, and we can infer that \gls{ris} {attains} 
similar or equal performance. Hence, we can conclude that the \gls{stcm} presents an advancement for {resource-efficient} sensing {design into ISAC} applications.

\section{Conclusions}\label{sec:VI}
In this paper, we provide a comprehensive theoretical study on using an \gls{stcm} as support for {sensing}, where the potential of detecting, localizing, and classifying {them} into distinct classes is assessed. The approach envisions future \gls{isac} applications since using \gls{stcm} has been proved to bring benefits also in communication \cite{Robin2021_fMIRS}. The results corroborate that using this kind of non-linear topology for intelligent metasurfaces can provide a subcentimeter to decimeter-level localization accuracy of \glspl{sp} and a method to effectively retrieve the \glspl{sp} position in local coordinates can be implemented with tradeoff only in processing by the \gls{bs} side; hence, integrating sensing and communication without degrading the communication service.
Although methods deploying linear intelligent metasurfaces had been proposed for sensing tasks, our proposed approach offers the advantage of using only narrowband pilot signals and the scenario being \textit{totally} static, not relying on Dopplers' processing of multiple transmitted symbols with high bandwidth \cite{kim2022}. The field of applications for the proposed method under the generic scenario investigated is vast, including electromagnetic field exposure aware communications, physical layer security, localization-based random access protocols, and precoding design based on environmental information, to name a few. Future works include investigating specific scenarios and developing algorithms to interface the metasurfaces' {topology} and the specific use cases.

\end{document}